\documentclass{iopjournal}

\usepackage{graphicx}
\usepackage{amsmath}
\usepackage{amssymb}
\usepackage{physics}
\usepackage{mathtools}
\usepackage[separate-uncertainty = true, print-unity-mantissa = false, per-mode = symbol]{siunitx}
\usepackage{hyperref}
\usepackage{xcolor}
\usepackage{booktabs}
\usepackage{parskip}
\usepackage{subcaption}
\usepackage[numbers,sort&compress]{natbib}
\usepackage{tikz} 
\usetikzlibrary{snakes,arrows,shapes}

\graphicspath{{figures/}{./}}

\definecolor{yellow1}{HTML}{D4A800}
\definecolor{yellow2}{HTML}{F0C800}
\definecolor{yellow3}{HTML}{FFF2A6}
\definecolor{mutedblue1}{HTML}{99A8CF}
\definecolor{mutedblue2}{HTML}{9EAAC9}
\definecolor{mutedblue3}{HTML}{A3ACC3}
\definecolor{gold1}{HTML}{E8C547}
\definecolor{gold2}{HTML}{DDC966}
\definecolor{gold3}{HTML}{D2CC85}

\newcommand{\SKR}{\mathrm{SKR}}
\newcommand{\DOP}{\mathrm{DOP}}

\begin{document}


\title{Deploying QKD on the fiber links of the National Quantum Communication Infrastructure in Sweden}

\author{
  Giulio Foletto$^{1,*,\orcid{0000-0002-9698-0909}}$,
  Mira Abu Yahia$^{1,\orcid{0009-0009-6203-1295}}$,
  Shreya Gouravaram Navalur$^{1,\orcid{0009-0003-4714-7528}}$,
  Carlos Brotons Montoro$^{1}$,
  Hilma Karlsson$^{1,\orcid{0009-0007-2107-1704}}$,
  Dena Karunianto Wibowo$^{1,\orcid{0000-0002-7240-4720}}$,
  Pascal Lefebvre$^{1,\orcid{0000-0001-8661-1440}}$,
  Katia Gallo$^{1,*,\orcid{0000-0001-7185-0457}}$
}

\affil{$^1$KTH Royal Institute of Technology, Stockholm, Sweden}

\affil{$^*$ Corresponding authors.}

\email{foletto@kth.se, gallo@kth.se}

\date{\today}

\begin{abstract}
  Quantum key distribution (QKD) is being deployed across Europe to evaluate its performance and practicality in real-world scenarios.
  When preparing this deployment, it is important to survey the properties of the fiber links that will be used for the quantum channel, cross-checking them with those of QKD devices, to ensure compatibility and optimal performance.
  We report here our studies from the Swedish chapter of this endeavor, the National Quantum Communication Infrastructure in Sweden.
  We focus on procedures to characterize the links and evaluate their suitability for QKD.
  We relate measurable properties to the key factors that affect the secret key rate using a simplified model that aims at the essence of QKD performance.
  Finally, we confirm the feasibility of QKD on one of our links by reporting a one-month long successful test.
\end{abstract}


\section{Introduction}
\label{sec:introduction}

Quantum key distribution (QKD) is a promising technology for secure communication, which allows the exchange of symmetric cryptographic keys with information-theoretic security based on the principles of quantum mechanics \cite{Wiesner1983,Bennett2014Rescan,Gisin2002,Scarani2009,Lo2014}.
Motivated also by the threat of the Shor algorithm on public-key cryptography \cite{Shor1997}, QKD has been moving steadily out of academic research laboratories and into the realm of practical implementations \cite{Xu2020,Pirandola2020,Chen2021,Chen2021a,Chen2025}.
In the EU, the European Quantum Communication Infrastructure (EuroQCI) initiative has been established in 2023 to begin the deployment of an interconnected quantum communication network across the continent, starting with national QKD infrastructures in almost all EU member states \cite{EuroQCI2023}.
During the first three year long phase, under the aegis of the Digital Europe Programme, fully functional networks have been demonstrated in several countries, with the plan of focusing on cross-border links in a subsequent phase.

The National Quantum Communication Infrastructure in Sweden (NQCIS) has been the Swedish chapter of this first phase of EuroQCI, and has been launched in 2023 by a consortium of research institutions (KTH -- Royal Institute of Technology, Stockholm University, Chalmers Technical University, Linköping University) and industry partners (Ericsson AB, Quantum Scopes AB, quCertify AB) \cite{NQCIS2023}.
Its main goal has been the procurement, deployment and evaluation of a network of commercial QKD devices connected by optical fiber links centered in Stockholm, playing the role of an open testbed for QKD in Sweden.
In addition to a number of research results, some in preparation, some already published \cite{Argillander2025,Karlsson2025,Clason2025,Spegel-Lexne2025,Almlof2025,Ahmadian2025,Ahmadian2025a}, NQCIS has had the effect of building Swedish expertise in the field of quantum communication, and of connecting stakeholders in academia and industry to prepare for future advances in practical QKD.

The initial setup of the network was challenging, and we encountered peculiar difficulties that are probably shared with most similar projects and with anyone first deploying a QKD network.
In particular, the simultaneous selection of fiber links and QKD devices proved critical, as the properties of the former had to be cross-referenced with the requirements of the latter.
To aid this process, we report here a detailed account of our experience in this regard, presenting measurement procedures and results for our network.
We relate the performance indicators of QKD to measurable properties of the fiber links, explaining which specifications of the QKD devices must be considered to make sure that key generation is possible.
We do so with a general performance model that only requires little information about the QKD devices, which can be used to evaluate the feasibility of QKD on any link.
Finally, we report a QKD test on one of our links, verifying not only in theory, but also experimentally, the applicability of QKD to our connections.

In Sec.~\ref{sec:model}, we give a pedagogical introduction to the performance of QKD in a simplified theoretical framework, highlighting the essential factors that limit key generation.
In Sec.~\ref{sec:links}, we present the physical properties of our specific links and the procedures used to assess them, connecting them to QKD performance.
In Sec.~\ref{sec:test}, we exemplify the use of our model to evaluate the feasibility of key generation, and compare this with actual results obtained from a commercial system.
Although the results are specific, the methodology is quite general and can be adapted to different protocols or links.
Hence, it can help others to select links and devices and make sure that requirements can be met in actual QKD deployments.

\section{Essential model of QKD performance}
\label{sec:model}

To directly relate measurable physical properties of the links to QKD performance, we extract from the literature \cite{Ma2005} an analytical model that expresses the secret key rate (SKR) as a function of as few as possible realistic parameters.
We focus on the efficient BB84 protocol, which is among the most well-known schemes for discrete-variable QKD, and can function as the archetype for many other protocols in this family \cite{Lo2005}.
This is a qubit protocol, i.e.\ one that uses two-level quantum systems, which are often implemented with light polarization or time-bin modes.
We consider an implementation that uses attenuated laser pulses to approximate single-photon states, and that relies on decoy states (three intensity levels: signal, decoy, and vacuum) to protect against photon-number-splitting attacks.

We work in a simplified scenario that trades realism for ease of exposition, aiming at the essence of QKD performance.
The first and foremost simplification is that we do not consider the effect of finite-size statistics and choose to work in the asymptotic limit of infinitely long keys.
In turn, this allows us to neglect some performance-degrading terms like the finite sifting factor and the Shannon inefficiency of error correction.
For the former, it is sufficient to assume that the users select the key-encoding basis and signal intensity almost all the times, which they can do with infinite keys.
For the latter, we use the fact that, in principle, error correction codes can approach the Shannon limit if applied to sufficiently long keys \cite{Tomamichel2017}.

In this framework, the SKR per transmitted pulse can be expressed as:
\begin{equation}
  \SKR = Q_1 \left(1 - h(E_1)\right)- Q_\mu h(E_\mu)
  \label{eq:skr}
\end{equation}
where $Q_1$ is the single-photon gain, i.e., the probability that a single-photon pulse is produced at the transmitter and causes a detection event at the receiver, $E_1$ is the phase error rate of these single-photon events, $Q_\mu$ is the overall signal gain, i.e., the probability that a signal pulse is produced and causes a detection event, and $E_\mu$ is the bit error rate of these events.
The function $h$ is the Shannon binary entropy, written as $h(x) = -x \log_2 x - (1-x) \log_2 (1-x)$ for $x \in (0,1)$ and extended via continuity to $h(0) = h(1) = 0$.

To express the four parameters $Q_1$, $E_1$, $Q_\mu$, and $E_\mu$, we consider threshold detectors that yield an event if they interact with a state containing at least one photon (after correction by their efficiency), and do not yield an event otherwise.
The qubit measurement is modeled as a beam splitter that transmits the incoming photon to one of two detectors with a non-zero probability of error.
The two exits of the beam splitter correspond to the two possible measurement outcomes, with one being correct and the other erroneous.

We note that in practice, while $Q_\mu$ and $E_\mu$ are directly observable at the beginning of post-processing, $Q_1$ and $E_1$ are not, and must be estimated from the measured quantities using the decoy-state method \cite{Hwang2003,Wang2005}.
However, in an ideal implementation in which the intensity of the decoy level is close to zero, the vacuum+weak scheme is known to approach ideality \cite{Ma2005}, therefore we are allowed to estimate $Q_1$ and $E_1$ using the physics of single-photon transmission too.

Given this, and setting the average intensity of the signal level to $\mu=1$ photon per pulse, the four parameters can be expressed as:
\begin{align}
  Q_1 &= e^{-1} \left(\eta + Y_0 - \eta Y_0\right), \label{eq:q1} \\
  E_1 &= \frac{q\eta + \frac12 Y_0}{\eta + Y_0 - \eta Y_0}, \label{eq:e1}\\
  Q_\mu &= 1 - e^{-\eta} (1 - Y_0), \label{eq:qmu}\\
  E_\mu &= \frac{q\left(1 - e^{-\eta}\right) + \frac12 Y_0}{1 - e^{-\eta} (1 - Y_0)}, \label{eq:emu}
\end{align}
where $\eta$ is the total transmittance, $Y_0$ is the probability that a vacuum pulse triggers a detection event, and $q$ represents the intrinsic quantum bit error rate, i.e. the probability that a photon prepared to encode a certain key bit is decoded as the other.
More accurately, $E_1$ and $E_\mu$ represent two different kinds of error rates: the phase-flip and bit-flip error rates, respectively.
However, in the asymptotic limit for the efficient BB84 protocol, the phase error rate in the key basis is estimated from the bit error rate in the control basis, so both terms descend from bit flipping probabilities, albeit in different bases.
In the absence of privileged basis (a valid approximation for polarization encoding, but less so for time bins), the error rates are expected to be the same, which justifies the use of a single parameter $q$ to describe both.

\begin{figure}
  \centering
  \begin{minipage}[t]{0.48\textwidth}
    \centering
    \includegraphics[width=\textwidth]{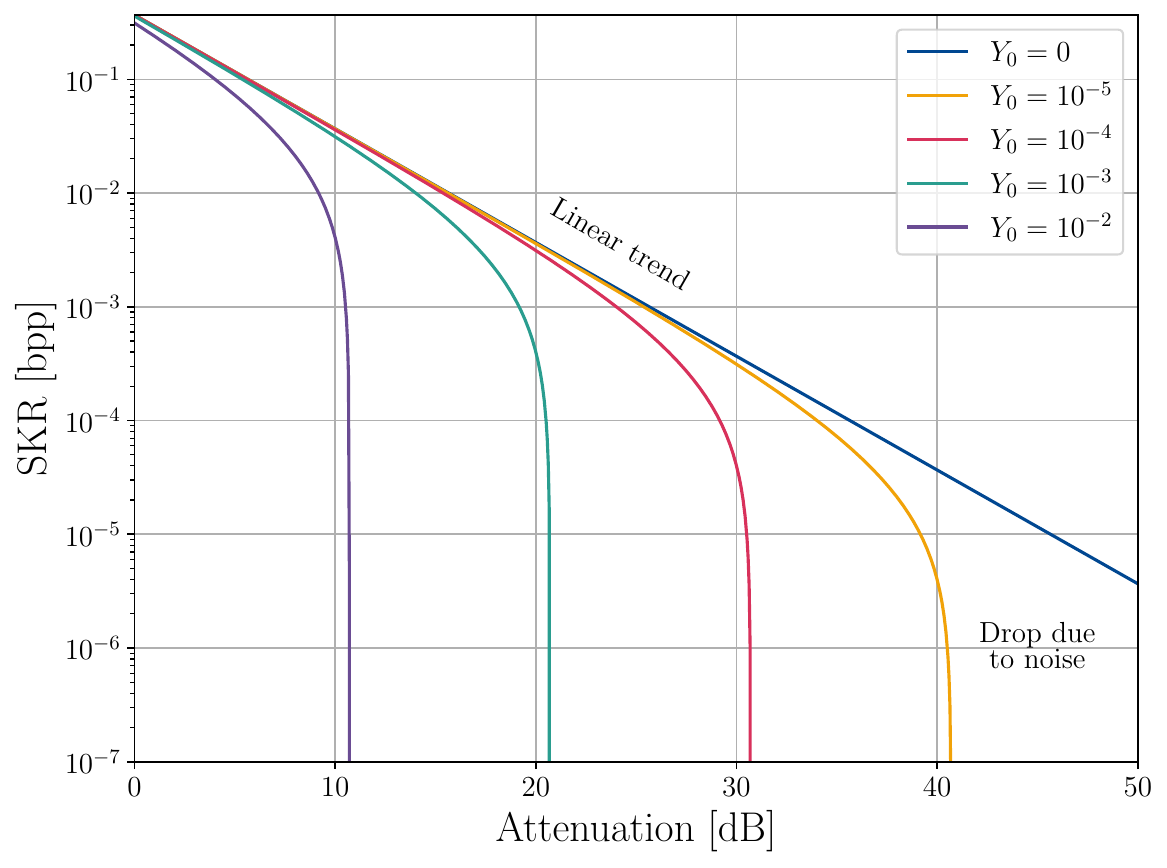}
    \caption{General scaling of the SKR as function of the attenuation.
    The \num{-1} slope indicates that $\SKR \propto \eta$.}
    \label{fig:skr_vs_attenuation_y0}
  \end{minipage}
  \hfill
  \begin{minipage}[t]{0.48\textwidth}
    \centering
    \includegraphics[width=\textwidth]{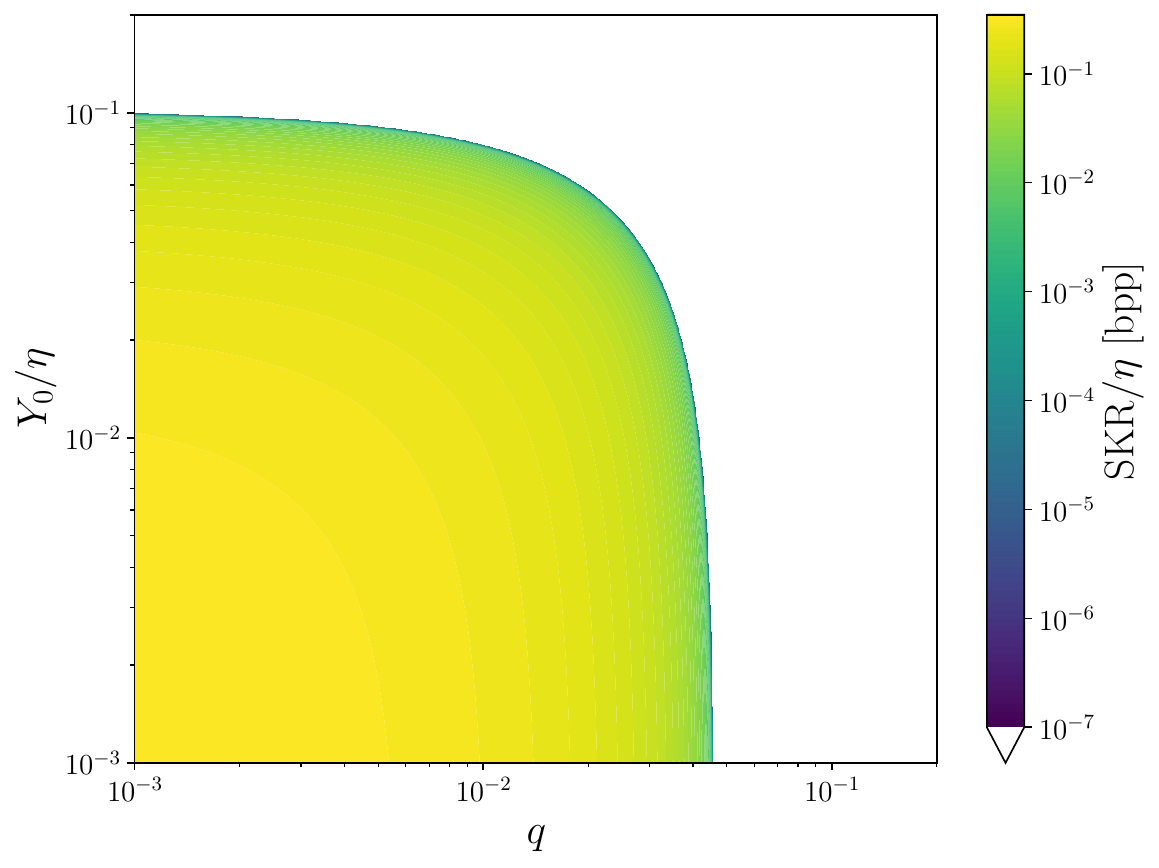}
    \caption{$\SKR/\eta$ against $q$ and $Y_0/\eta$, for fixed $\eta=\num{1e-3}$. The white region corresponds to $\SKR=0$.}
    \label{fig:skr_contour_vs_y0_q}
  \end{minipage}
\end{figure}

We have reduced the model to depend on only three parameters: $\eta$, $Y_0$, and $q$.
To provide a visual intuition of the effect of these parameters on the SKR, we plot in Fig.~\ref{fig:skr_vs_attenuation_y0} the SKR as a function of the attenuation $\eta_{\text{dB}} = -10\log_{10} \eta$ (expressed in decibel) for different values of $Y_0$, and setting $q=0$.
The \num{-1} slope of the first part of the curves is a signature of the direct proportionality between the SKR and the transmittance, which is also evident from Eqs.~\eqref{eq:q1} and \eqref{eq:emu} which are dominated by a term that is linear in $\eta$.
This is a fundamental limit: a lost photon cannot contribute to the generation of the key \cite{Pirandola2017}.
It can only be overcome with substantial modifications to the protocol, such as twin-field QKD \cite{Lucamarini2018,Liu2023}, or the use of quantum repeaters, which are still in the research phase and far from practical availability \cite{Sangouard2011,Munro2015,Azuma2023}.
The quick drop to zero of the SKR at high attenuations is due to the fact that the background noise $Y_0$ becomes dominant with respect to the detected signal, so that the QBER exceeds the threshold for secure key generation.
The minimum tolerable transmittance is linearly proportional to $Y_0$, as can be seen by the fact that the position of the drop, which represents the maximum tolerable attenuation, moves to the left by \SI{10}{\deci\bel} as $Y_0$ increases by a factor of \num{10}.

Using the linear relation between $\eta$ and $\SKR$, we plot in Fig.~\ref{fig:skr_contour_vs_y0_q} the computed value of $\SKR/\eta$ as a function of both $q$ and $Y_0/\eta$, for fixed $\eta=\num{1e-3}$.
Although the exact values depend on this choice of $\eta$, the boundary that gives $\SKR=0$ remains approximately the same for any $\eta \ll 1$.
We can see that the $\SKR$ decreases only slowly if $q$ and $Y_0/\eta$ stay below \SI{1}{\percent}, but drops quickly to zero when they get close to their maximum tolerable values, which are approximately \SI{4.6}{\percent} and \num{0.1}, respectively.
These values are due to our choice of decoy protocol and signal intensity, and should be recalculated on a case-by-case basis (for instance if the intensity levels change).
However, the methodology is general and tells us that overoptimizing $q$ or $Y_0$ yields diminishing returns, and the focus should be on keeping them much smaller than a computable threshold.

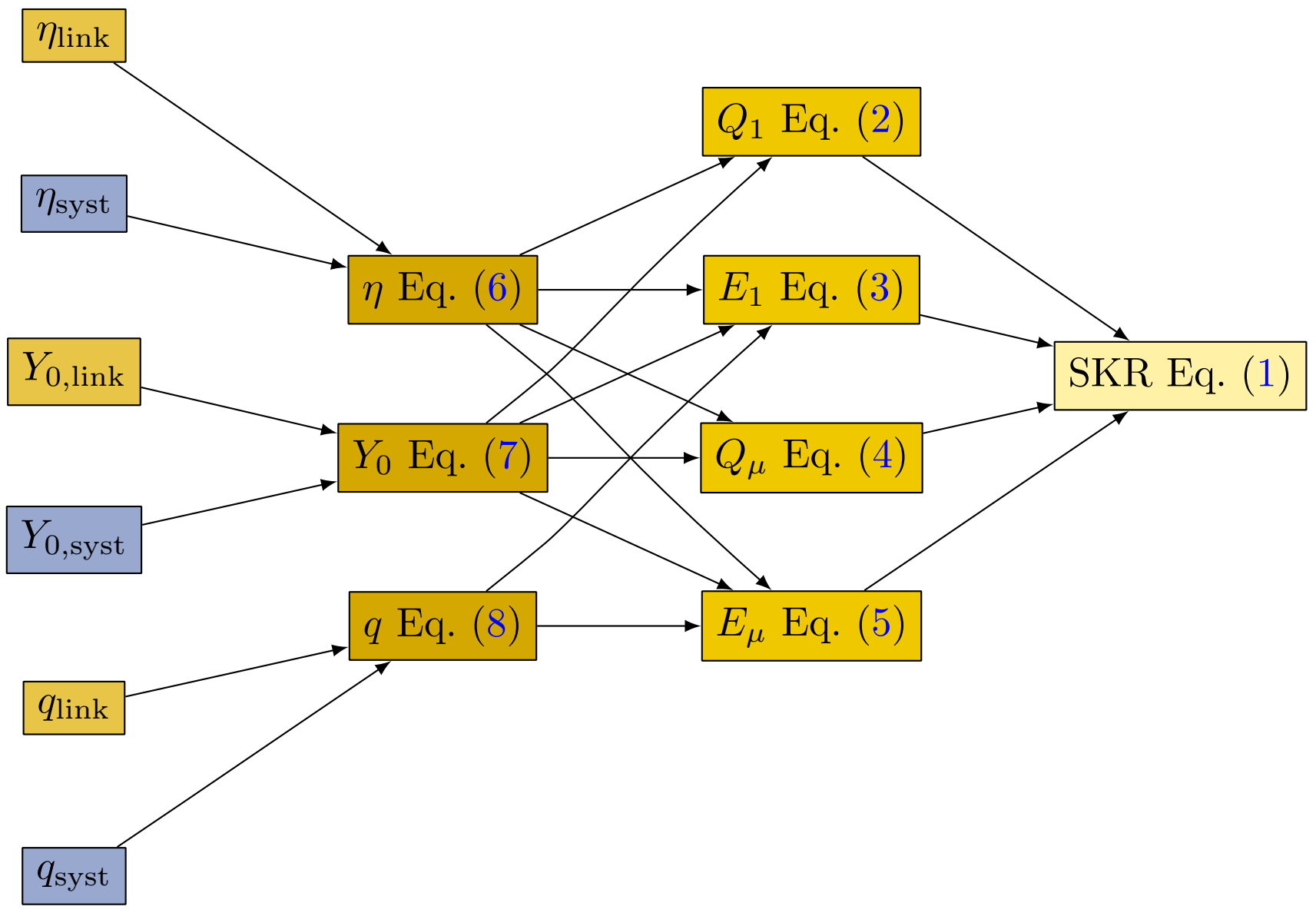
\begin{figure}
  \centering
  \begin{tikzpicture}[>=latex,line join=bevel,]
\begin{scope}
  \pgfsetstrokecolor{black}
  \definecolor{strokecol}{rgb}{1.0,1.0,1.0};
  \pgfsetstrokecolor{strokecol}
  \definecolor{fillcol}{rgb}{1.0,1.0,1.0};
  \pgfsetfillcolor{fillcol}
  \filldraw (0.0bp,0.0bp) -- (0.0bp,241.0bp) -- (324.0bp,241.0bp) -- (324.0bp,0.0bp) -- cycle;
\end{scope}
\begin{scope}
  \pgfsetstrokecolor{black}
  \definecolor{strokecol}{rgb}{1.0,1.0,1.0};
  \pgfsetstrokecolor{strokecol}
  \definecolor{fillcol}{rgb}{1.0,1.0,1.0};
  \pgfsetfillcolor{fillcol}
  \filldraw (0.0bp,0.0bp) -- (0.0bp,241.0bp) -- (324.0bp,241.0bp) -- (324.0bp,0.0bp) -- cycle;
\end{scope}
\begin{scope}
  \pgfsetstrokecolor{black}
  \definecolor{strokecol}{rgb}{1.0,1.0,1.0};
  \pgfsetstrokecolor{strokecol}
  \definecolor{fillcol}{rgb}{1.0,1.0,1.0};
  \pgfsetfillcolor{fillcol}
  \filldraw (0.0bp,0.0bp) -- (0.0bp,241.0bp) -- (324.0bp,241.0bp) -- (324.0bp,0.0bp) -- cycle;
\end{scope}
  \node (q1) at (207.0bp,202.0bp) [draw,fill=yellow2,rectangle] {$Q_1$ Eq.~\eqref{eq:q1}};
  \node (e1) at (207.0bp,161.0bp) [draw,fill=yellow2,rectangle] {$E_1$ Eq.~\eqref{eq:e1}};
  \node (qmu) at (207.0bp,120.0bp) [draw,fill=yellow2,rectangle] {$Q_\mu$ Eq.~\eqref{eq:qmu}};
  \node (emu) at (207.0bp,79.0bp) [draw,fill=yellow2,rectangle] {$E_\mu$ Eq.~\eqref{eq:emu}};
  \node (eta) at (117.0bp,161.0bp) [draw,fill=yellow1,rectangle] {$\eta$ Eq.~\eqref{eq:eta}};
  \node (y0) at (117.0bp,120.0bp) [draw,fill=yellow1,rectangle] {$Y_0$ Eq.~\eqref{eq:y0}};
  \node (q) at (117.0bp,79.0bp) [draw,fill=yellow1,rectangle] {$q$ Eq.~\eqref{eq:q}};
  \node (etal) at (27.0bp,223.0bp) [draw,fill=gold1,rectangle] {$\eta_{\mathrm{link}}$};
  \node (etas) at (27.0bp,182.0bp) [draw,fill=mutedblue1,rectangle] {$\eta_{\mathrm{syst}}$};
  \node (y0l) at (27.0bp,141.0bp) [draw,fill=gold1,rectangle] {$Y_{0, \mathrm{link}}$};
  \node (y0s) at (27.0bp,100.0bp) [draw,fill=mutedblue1,rectangle] {$Y_{0, \mathrm{syst}}$};
  \node (ql) at (27.0bp,59.0bp) [draw,fill=gold1,rectangle] {$q_{\mathrm{link}}$};
  \node (qs) at (27.0bp,18.0bp) [draw,fill=mutedblue1,rectangle] {$q_{\mathrm{syst}}$};
  \node (skr) at (297.0bp,140.0bp) [draw,fill=yellow3,rectangle] {$\SKR$ Eq.~\eqref{eq:skr}};
  \draw [->] (q1) ..controls (242.27bp,177.86bp) and (251.66bp,171.24bp)  .. (skr);
  \draw [->] (e1) ..controls (241.97bp,152.89bp) and (250.38bp,150.89bp)  .. (skr);
  \draw [->] (qmu) ..controls (241.97bp,127.72bp) and (250.38bp,129.63bp)  .. (skr);
  \draw [->] (emu) ..controls (242.58bp,102.97bp) and (251.73bp,109.31bp)  .. (skr);
  \draw [->] (eta) ..controls (152.23bp,176.95bp) and (160.96bp,181.01bp)  .. (q1);
  \draw [->] (eta) ..controls (151.89bp,161.0bp) and (160.18bp,161.0bp)  .. (e1);
  \draw [->] (eta) ..controls (152.23bp,145.05bp) and (160.96bp,140.99bp)  .. (qmu);
  \draw [->] (eta) ..controls (141.37bp,141.45bp) and (142.71bp,140.22bp)  .. (144.0bp,139.0bp) .. controls (158.4bp,125.39bp) and (162.12bp,118.99bp)  .. (emu);
  \draw [->] (y0) ..controls (141.37bp,139.55bp) and (142.71bp,140.78bp)  .. (144.0bp,142.0bp) .. controls (158.4bp,155.61bp) and (162.12bp,162.01bp)  .. (q1);
  \draw [->] (y0) ..controls (152.23bp,135.95bp) and (160.96bp,140.01bp)  .. (e1);
  \draw [->] (y0) ..controls (151.89bp,120.0bp) and (160.18bp,120.0bp)  .. (qmu);
  \draw [->] (y0) ..controls (152.23bp,104.05bp) and (160.96bp,99.986bp)  .. (emu);
  \draw [->] (q) ..controls (141.37bp,98.546bp) and (142.71bp,99.781bp)  .. (144.0bp,101.0bp) .. controls (158.4bp,114.61bp) and (162.12bp,121.01bp)  .. (e1);
  \draw [->] (q) ..controls (151.89bp,79.0bp) and (160.18bp,79.0bp)  .. (emu);
  \draw [->] (etal) ..controls (62.267bp,198.86bp) and (71.655bp,192.24bp)  .. (eta);
  \draw [->] (etas) ..controls (61.973bp,173.89bp) and (70.376bp,171.89bp)  .. (eta);
  \draw [->] (y0l) ..controls (61.973bp,132.89bp) and (70.376bp,130.89bp)  .. (y0);
  \draw [->] (y0s) ..controls (61.973bp,107.72bp) and (70.376bp,109.63bp)  .. (y0);
  \draw [->] (ql) ..controls (61.973bp,66.721bp) and (70.376bp,68.631bp)  .. (q);
  \draw [->] (qs) ..controls (62.582bp,41.972bp) and (71.733bp,48.315bp)  .. (q);
\end{tikzpicture}
  \caption{Scheme of the simplified model for the performance of QKD.}
  \label{fig:graph_analysis}
\end{figure}

The three parameters $\eta$, $Y_0$, and $q$ depend not only on the link, but also on the QKD system.
Specifically, $\eta$ contains the link contribution, but also all losses at the receiver, including the efficiency of the single-photon detectors.
$Y_0$ contains the background rate coming from the link, but also the dark count rate of the detectors.
$q$ depends on the transformation applied to the quantum state by the link, but also on the quality of the preparation and of the measurement.
To combine these contributions, we can write:
\begin{align}
  \eta &= \eta_{\mathrm{link}}\cdot\eta_{\mathrm{syst}}\,,\label{eq:eta}\\
  Y_0 &= Y_{0,\mathrm{link}}+Y_{0,\mathrm{syst}}\,,\label{eq:y0}\\
  q &= q_{\mathrm{link}} + q_{\mathrm{syst}} -2q_{\mathrm{link}}q_{\mathrm{syst}}\,,\label{eq:q}
\end{align}
so that the model takes the graph form shown in Fig.~\ref{fig:graph_analysis}.

We note that there are several simplifications here in addition to those mentioned at the beginning of this section.
First, the distinction between link and system contributions is not always clear-cut, because QKD devices can employ countermeasures to mitigate the effect of link imperfections, such as polarization compensation or spectral filtering (which we will see in detail in Sec.~\ref{sec:links}).
Moreover, in practice there is often a non-trivial interplay between $\eta$, $Y_0$, and $q$; for example, a smaller $\eta$ or higher $Y_0$ can impair the functioning of compensation mechanisms, leading to a higher $q$.
Finally, Eq.~\eqref{eq:q} is an approximation that assumes that the two contributions to the QBER act as independent qubit depolarizing channels, whereas the real action can be much more complex.
However, this scheme is useful to guide the deployment process, highlighting that link and system parameters must be cross-checked and, if possible, optimized to make sure that QKD can operate to an acceptable performance level.
In the next section, we will see how measurable parameters can lead to the quantities on the left side of Fig.~\ref{fig:graph_analysis}, which then leads us directly to the SKR.

\section{The network and its properties}
\label{sec:links}

\begin{figure}
  \centering
  \includegraphics[width=0.5\textwidth]{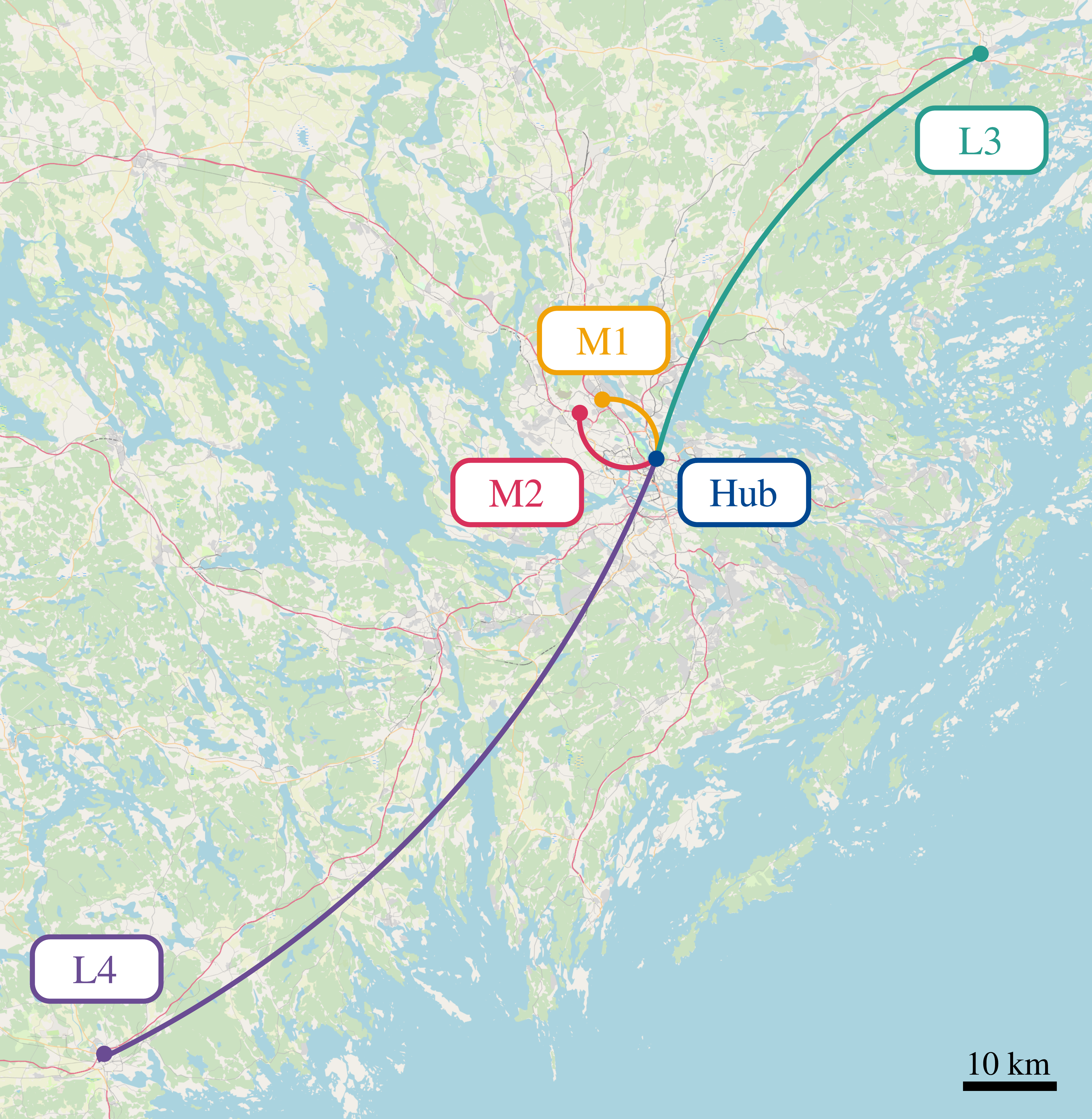}
  \caption{Map of the subset of the NQCIS network considered in this work.
  Copyright on base map data: OpenStreetMap contributors.}
  \label{fig:map}
\end{figure}

Our analysis focuses on four representative links used by the QKD network deployed by NQCIS.
The topology of this subset of the network is that of a star, with the main node, functioning as entry point and hub, being operated by KTH and located at the Albanova University Center in Stockholm.
As done in Fig.~\ref{fig:map}, we will refer to the four links using the labels M1, M2, L3, and L4.
The letters indicate that the former two links are part of a metropolitan network, connecting the hub to two nodes in Kista operated by Ericsson AB and the Research Institute of Sweden, whereas the latter two are part of a long-distance network, extending to Norrtälje to the north-east and Nyköping to the south-west.
Each link consists of two parallel strands of single-mode optical fiber, rented exclusively for NQCIS.

We now go through the physical properties of these links, and explain how they can be measured and how they affect the performance of QKD.
We start from the length, which influences all properties, and then focus on finding $\eta_{\mathrm{link}}$, $Y_{0,\mathrm{link}}$, and $q_{\mathrm{link}}$ from the attenuation, background noise, and polarization drift of the fiber.
The focus on polarization is more relevant for polarization-encoded QKD, but should not be dismissed for other encodings which still can see their performance affected by polarization effects.

\subsection{Fiber length}
\label{sec:length}

The most basic feature of an optical-fiber link is its length.
Its impact on the performance of QKD is almost always indirect: all other properties of the fiber depend on length, and QKD is affected by them.
Therefore, the length of a fiber link can give an initial idea of its suitability for QKD, although, as we will see, it is not sufficient for a complete assessment.

Fiber length can be measured with several methods.
The most common employs an optical time-domain reflectometer (OTDR), which sends a series of light pulses into the fiber and measures the time it takes for the reflected light to return, allowing to calculate the length from the known speed of light in the material.
This technique is widely used in the telecommunications industry due to its practicality: OTDRs are commercially available as packaged and portable devices, and can be used in the field with minimal training.
Moreover, as we will see in Sec.~\ref{sec:attenuation}, the OTDR trace can provide not only the length of the fiber, but also information about the location and magnitude of any losses or reflections along its path, which can be useful for diagnosing issues with the link.

\begin{table}
  \centering
  \caption{Measured lengths and attenuations (at \SI{1550}{\nano\meter}) of the fiber links in the NQCIS network.}
  \begin{tabular}{ccccc}
    \toprule
    Link & Length [\unit{\kilo\meter}] & Tortuosity & Attenuation [\unit{\dB}] & Eff.\ att.\ coeff.\ [\unit{\dB\per\kilo\meter}] \\
    \midrule
    M1--A & \num{18.50 \pm 0.01} & \num{2.29} & \num{9.18 \pm 0.03} & \num{0.496 \pm 0.002} \\
    M1--B & \num{18.50 \pm 0.01} & \num{2.29} & \num{8.59 \pm 0.03} & \num{0.464 \pm 0.002} \\
    M2--A & \num{19.73 \pm 0.01} & \num{2.37} & \num{7.49 \pm 0.03} & \num{0.380 \pm 0.002} \\
    M2--B & \num{19.73 \pm 0.01} & \num{2.37} & \num{7.16 \pm 0.03} & \num{0.363 \pm 0.002} \\
    L3--A & \num{80.85 \pm 0.02} & \num{1.39} & \num{19.32 \pm 0.04} & \num{0.239 \pm 0.002} \\
    L3--B & \num{80.86 \pm 0.02} & \num{1.39} & \num{20.63 \pm 0.04} & \num{0.255 \pm 0.002} \\
    L4--A & \num{163.63 \pm 0.03} & \num{1.60} & \num{36.8 \pm 0.1} & \num{0.225 \pm 0.002} \\
    L4--B & \num{163.63 \pm 0.03} & \num{1.60} & \num{38.0 \pm 0.1} & \num{0.232 \pm 0.002} \\
    \bottomrule
  \end{tabular}
  \label{tab:lengths_attenuations}
\end{table}

We measured the length of the links with a commercial OTDR device (Anritsu MW9076B), and report the results, together with attenuation values (see Sec.~\ref{sec:attenuation}), in Tab.~\ref{tab:lengths_attenuations}.
Since the L4 link is longer than the range of our OTDR, we report the value recorded by the fiber supplier.
The uncertainties are those provided by the device.
Notably, the NQCIS network consists of links in three different length classes ($\sim \num{20}, \num{80}, \SI{160}{\kilo\meter}$), which allows a comprehensive analysis of fiber properties.
The two parallel strands of each link, labeled A and B, have compatible lengths, within one standard deviation.

We also report the ratio between the measured length and the (geodesic) distance between the two endpoints, which we call tortuosity.
All metropolitan links show high values well above \num{2}, and a maximum of \num{2.37} is reached for the M2 link.
This is compatible with the fact that fiber networks in this area are likely to be composed of many short segments, to offer easy reconfigurability and serve a high density of users.
The tortuosity decreases for longer links, which travel through less populated regions and are more likely to follow a straight path along highways or other major routes.
This variability means that, when planning a QKD deployment, it is not sufficient to consider only the distance between the nodes, but it is necessary to interact with fiber providers to select connections with the appropriate length and properties.

\subsection{Attenuation in optical fibers}
\label{sec:attenuation}

The primary factor driving attenuation in optical-fiber links is the physics of the fiber material, which inevitably absorbs and scatters light.
Because this phenomenon happens uniformly along the fiber, it is quantified by an attenuation coefficient in \unit{\dB\per\kilo\meter}, which is a measure of the loss per unit length.
For typical single-mode fibers used in telecommunications, this coefficient is minimized in the \SIrange{1500}{1600}{\nano\meter} wavelength range, where it usually takes value around \SI{0.2}{\dB\per\kilo\meter}.
This leads to an exponential decrease of transmittance (or linear increase of attenuation), with the length of the fiber.
However, localized imperfections, such as splices, connectors, or bends, can cause additional losses that are not directly related to length.
This makes the effective attenuation coefficient usually much higher than the nominal value.

\begin{figure}
  \centering
  \includegraphics[width=0.5\textwidth]{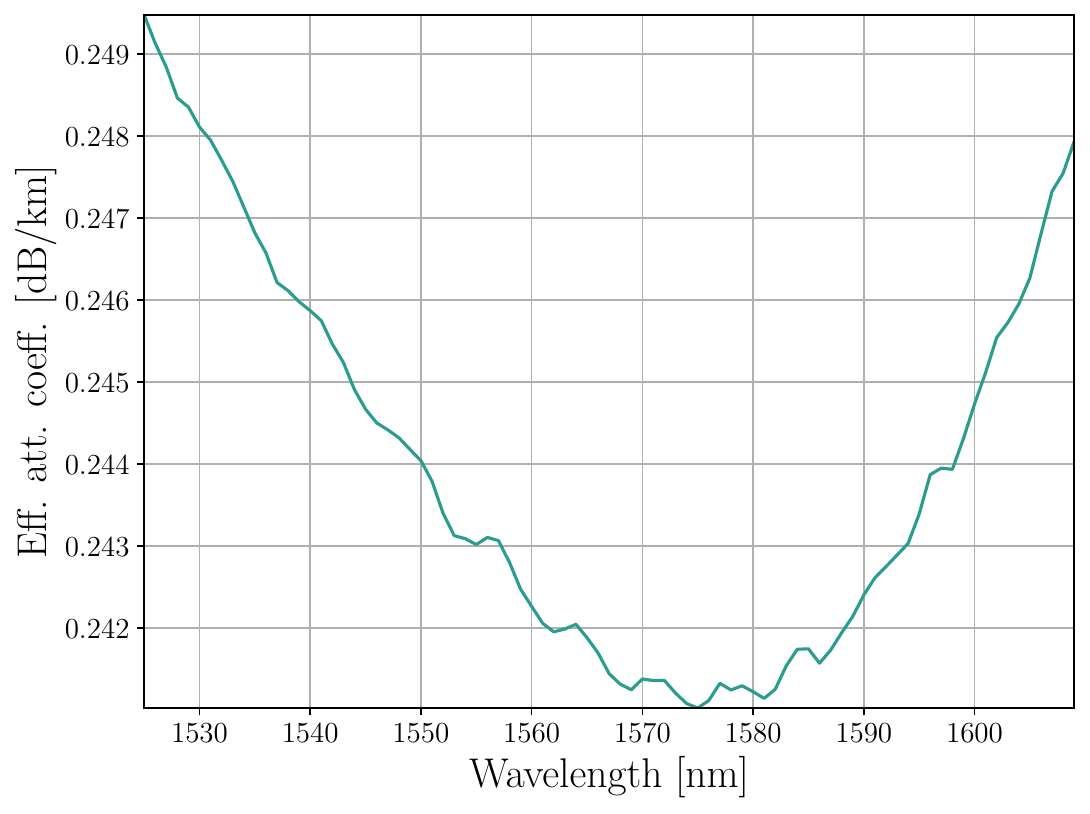}
  \caption{Attenuation coefficient as a function of wavelength for the looped L3 link. The minimum occurs around \SI{1575}{\nano\meter}.}
  \label{fig:att_vs_wl}
\end{figure}

Attenuation can be measured by sending light through a fiber link and recording the optical power at the input and output ends.
To study the behavior of the attenuation coefficient in a broad range of wavelengths, we use a tunable laser (IDPhotonics Cobrite DX2-LC) on the looped L3 link, i.e. with a patch connecting the two strands at the remote end, and measure the optical power with a power meter (Thorlabs S155C).
The results, shown in Fig.~\ref{fig:att_vs_wl}, indicate the presence of a minimum around the wavelength of \SI{1575}{\nano\meter}, where the effective attenuation coefficient is approximately \SI{0.241}{\dB\per\kilo\meter}, whereas the maximum occurs at the edges of the studied range, with a value of around \SI{0.249}{\dB\per\kilo\meter} at \SI{1525}{\nano\meter}.

Although this \SI{3}{\percent} variation may seem small, it can have a significant impact on long-distance links, considering that transmittance scales exponentially with both length and attenuation coefficient.
For instance, for the full length of the looped L3 link of approximately \SI{162}{\kilo\meter}, this change would correspond to \SI{1.3}{\dB} of additional loss, leading to \SI{26}{\percent} lower $\SKR$ even without considering the effect of the reduced $Y_0/\eta$ ratio.
However, as we shall see later, because noise also depends on wavelength, basing the selection on attenuation alone may not yield the optimal performance.
Moreover, the position of the minimum on the wavelength axis can change from link to link, probably because of variations in the fiber manufacturing process and the specific installation conditions.
For instance we repeated the test on a fiber spool and found a decreasing trend of the attenuation coefficient with increasing wavelength, suggesting a minimum beyond \SI{1610}{\nano\meter}.
For this and for practicality, we focus the rest of our analysis on the \SI{1550}{\nano\meter} wavelength, which is commonly used in telecommunications and offers a good compromise between low attenuation and availability of components.
At this wavelength, the effective attenuation coefficient in Fig.~\ref{fig:att_vs_wl} is approximately \SI{0.244}{\dB\per\kilo\meter}.

\begin{figure}
  \centering
  \begin{minipage}[t]{0.48\textwidth}
    \centering
    \includegraphics[width=\textwidth]{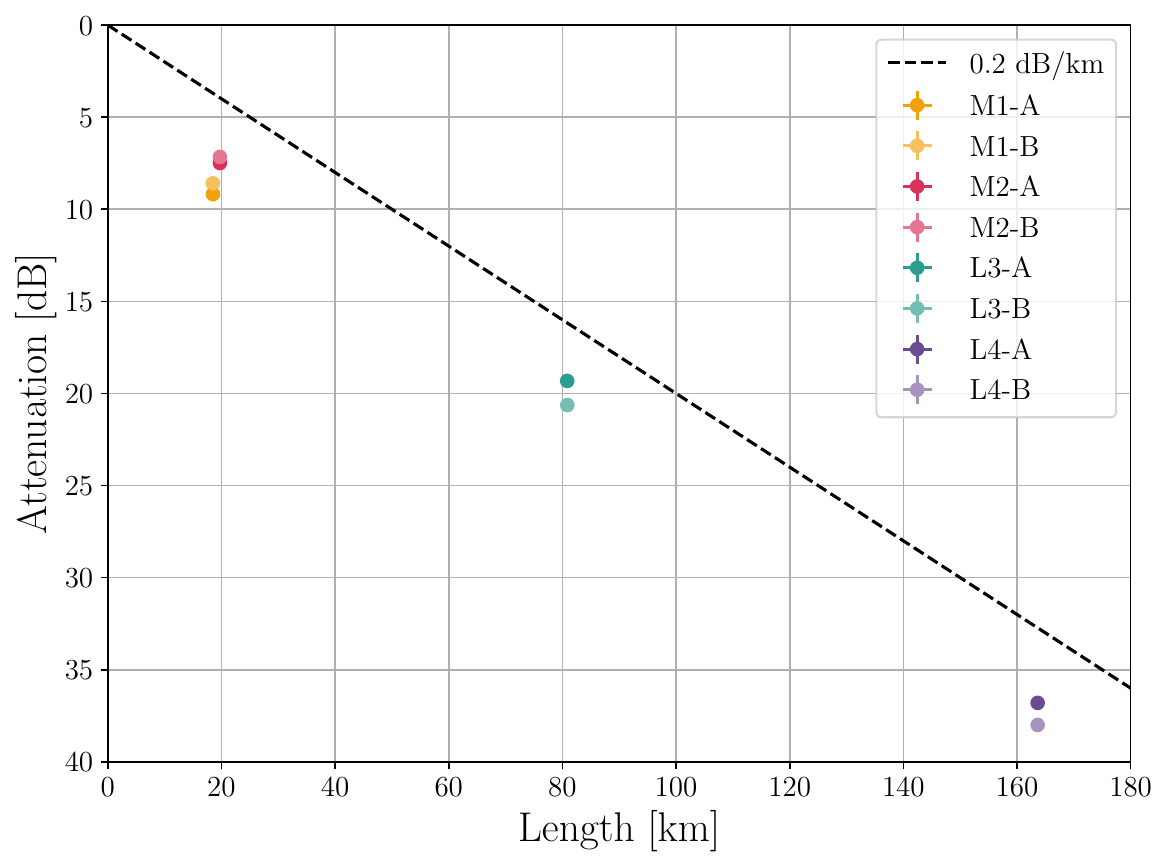}
    \caption{Measured attenuation values and general scaling against length.}
    \label{fig:attenuation_vs_length}
  \end{minipage}
  \hfill
  \begin{minipage}[t]{0.48\textwidth}
    \centering
    \includegraphics[width=\textwidth]{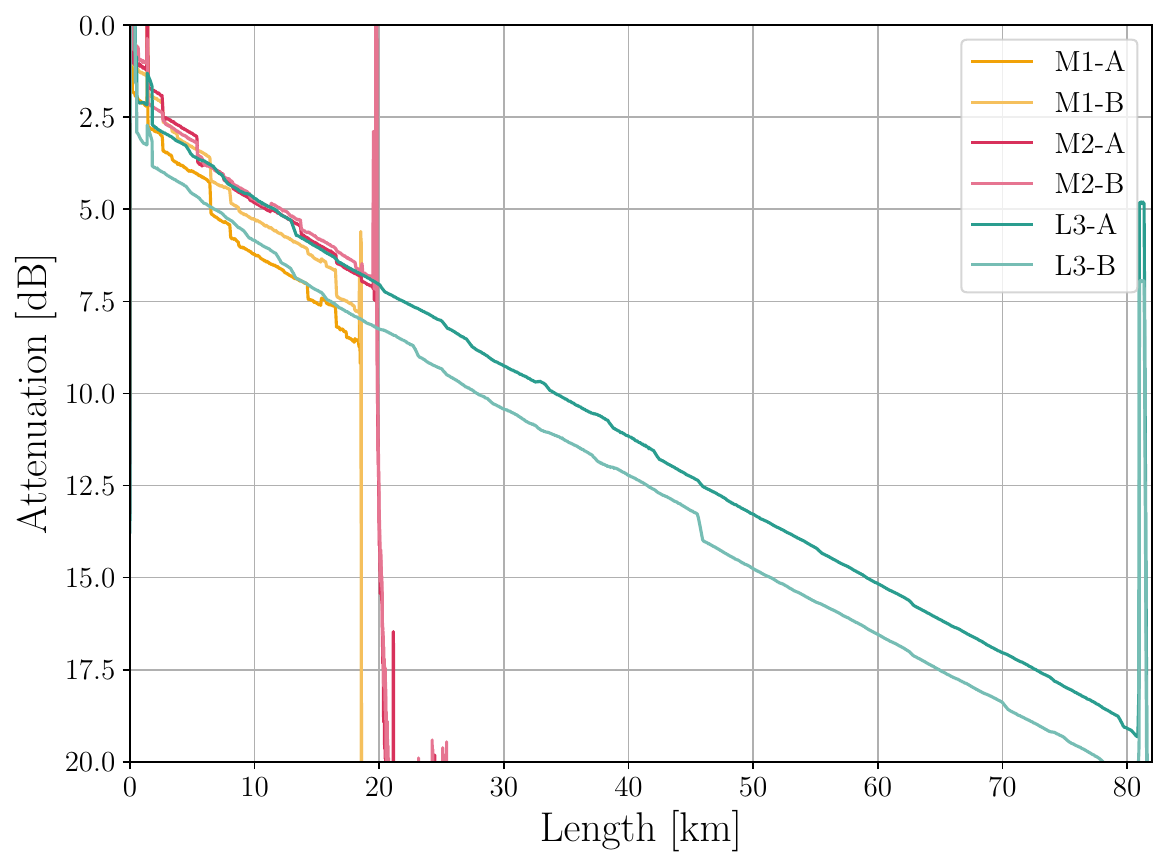}
    \caption{OTDR traces for the M1, M2, and L3 links.}
    \label{fig:otdr_plot_multi}
  \end{minipage}
\end{figure}

We characterize all our links strand-by-strand (without any loop) with the same setup as before.
The measured attenuations at \SI{1550}{\nano\meter} are provided in Tab.~\ref{tab:lengths_attenuations} and Fig.~\ref{fig:attenuation_vs_length}, which shows also the comparison with the nominal \SI{0.2}{\dB\per\kilo\meter} coefficient that is considered the standard value for typical single-mode fibers.

The variability of the values, ranging from \num{7} to \SI{38}{\dB}, gives the opportunity to study QKD in various conditions, from relatively easy to very challenging.
The two strands in each link show a non-negligible difference in attenuation.
For example, there is more than \SI{1}{\dB} difference between the two strands of each L link, but also \SI{0.5}{\dB} for the much shorter M1 link.
This highlights the fact that even fibers following the same route can exhibit significant differences in attenuation due to variations in the quality of splices, connectors, and other localized imperfections.

As for the tortuosity, the metropolitan links show a higher attenuation coefficient, with the maximum, \SI{0.496\pm0.002}{\dB\per\kilo\meter}, obtained on the M1-A link.
This agrees with the previous observation that metropolitan links are composed of many short segments.
Each connector or splice increases the attenuation, and the cumulative effect of many such imperfections can be significant.

To show the impact of these localized imperfections, we plot the OTDR traces for the three links that our device can measure in Fig.~\ref{fig:otdr_plot_multi}.
These traces show a higher density of imperfections (jumps) for the metropolitan links, and for the initial part of the L3 link, which is closer to the city.
They also show a drop around the \SI{45}{\kilo\meter} mark for the L3-B link, which justifies a good portion of the difference in attenuation with respect to its sister strand.

All of this indicates that the precise network topology and individual fiber conditions are both relevant, and predicting the attenuation based only on the length of the link can be misleading.
Furthermore, during the lifetime of a fiber, new splices can be added to repair damage, which can further exacerbate deviations from expected performance.
Accurate and repeated characterizations of attenuation are therefore an important step of network design and maintenance.

\subsection{Background photon noise in dark fibers}
\label{sec:noise}

Even fibers that are nominally dark, such as those in the NQCIS network, can carry a non-negligible amount of background photons, which reduce QKD performance by increasing $Y_0$.
Physically, this noise can arise from various sources, such as ambient light in sections where the fiber is exposed, optical signals from other strands leaking via evanescent coupling, or even thermal radiation.

QKD devices can use two main countermeasures against link-induced background noise, which are temporal and spectral filtering.
If the signal pulses are much shorter in time than the repetition period of the source, and detectors are sufficiently precise in the temporal determination of events, it is possible to accept events that occur only in a narrow time window around the expected arrival time of the signal, and reject all others.
Similarly, if the signal pulses are narrow enough in wavelength, the application of a spectral filter at the detector can substantially reduce the value of $Y_0$ for the same incoming background light.

A relevant difference between the two techniques is that temporal filtering requires synchronization with the transmitter, and as such is usually implemented by the QKD-device manufacturer, whereas spectral filtering can be implemented by the user with off-the-shelf components, requiring only knowledge of the signal wavelength.
Therefore, at link characterization time, it is useful to measure not only the total background photon rate, but also what remains of it after using a band-pass filter at the right wavelength.
We can write:
\begin{align}
  R_{\mathrm{bkg}} &= \int_0^{\infty} b_\lambda\, F_\lambda \, d\lambda\,, \label{eq:rbkg}\\
  Y_{0,\mathrm{link}} &= R_{\mathrm{bkg}} \cdot \Delta t\,,\label{eq:y0link}
\end{align}
where $R_{\mathrm{bkg}}$ is the measurable background rate after filtering, $b_\lambda$ is the background photon rate density in \unit{\hertz\per\nano\meter}, $F_\lambda$ is the transfer function of the spectral filter, and $\Delta t$ is the width of the temporal acceptance window implemented by the device.
The device can further reduce the effect by using a narrower spectral filter, up to the bandwidth of the transmitter laser, but $R_{\mathrm{bkg}}$ provides a baseline for the background noise that is available to the users without reliance on detailed specifications of the device and that is sufficient for feasibility assessment.

We note that if $F_\lambda$ has a non-negligible attenuation at the signal wavelength, this should be accounted for and combined with $\eta_{\mathrm{link}}$.
Moreover, the action of $\Delta t$ is equivalent to that of Eq.~\eqref{eq:y0link} also when applied to the dark count rate of the detectors used by the QKD devices:
\begin{equation}
  Y_{0,\mathrm{syst}} = R_{\mathrm{dark}} \cdot \Delta t\,.
  \label{eq:y0syst}
\end{equation}

\begin{figure}
  \centering
  \includegraphics[width=0.5\textwidth]{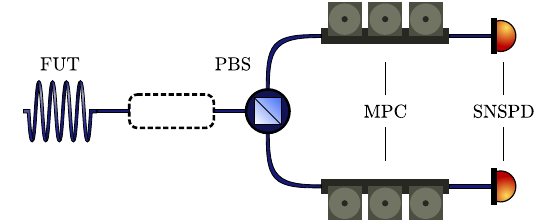}
  \caption{Scheme of the measurement setup for the background light in deployed fiber links.
  FUT: fiber under test, PBS: in-fiber polarizing beam splitter, MPC: manual polarization controller, SNSPD: superconducting nanowire single photon detector.
  The dashed box represents a component that changes in the different measurements: a direct fiber connection for the measurement of total background rate, a tunable wavelength filter for the measurement of the noise spectrum, and another MPC for the measurement of the polarization extinction ratio.}
  \label{fig:background_noise_setup}
\end{figure}

In-field measurements of the background photon rate are challenging, because the typical telecom equipment used for fiber characterization is not designed to detect light at such low powers.
However, portable single-photon avalanche diodes (SPADs) might make this easier in the near future.
We measured the background photon noise using superconducting nanowire single-photon detectors (SNSPDs, SingleQuantum EOS) installed in the hub, which feature very low dark count rates and high quantum efficiency.
Due to the polarization-dependent efficiency of our detectors, we measured the noise at the same time in two polarization components, separated with an in-fiber polarizing beam splitter (Thorlabs PBC1550SM-FC).
The two outputs are each aligned with manual polarization controllers (MPC, Thorlabs FPC561) to the best polarization axis for the SNSPDs, and sent to two different detectors with known efficiency, so that the total noise can be estimated by summing the two contributions.
A scheme of the setup is shown in Fig.~\ref{fig:background_noise_setup}.

\begin{table}
  \centering
  \caption{Properties of the background photon noise in the fiber links of the NQCIS network.
  The rate columns report sample averages and standard deviations.}
  \begin{tabular}{ccccc}
    \toprule
    Link & Unfilt.\ rate [\unit{\kilo\hertz}] & Filt.\ rate [\unit{\kilo\hertz}] & Extinction ratio [\unit{\dB}] & DOP [\unit{\percent}] \\
    \midrule
    M1--A & $\num{19.5} \pm \num{1.3}$ & $\num{0.69} \pm \num{0.05}$ & \num{14.5} & \num{19} \\
    M1--B & $\num{12.4} \pm \num{0.6}$ & $\num{0.39} \pm \num{0.02}$ & \num{15.0} & \num{18} \\
    M2--A & $\num{13.3} \pm \num{0.7}$ & $\num{0.008} \pm \num{0.001}$ & \num{32.1} & \num{32} \\
    M2--B & $\num{6.9} \pm \num{0.3}$ & $\num{0.004} \pm \num{0.001}$ & \num{32.7} & \num{46} \\
    L3--A & $\num{476} \pm \num{49}$ & $\num{0.9} \pm \num{0.1}$ & \num{27.1} & \num{29} \\
    L3--B & $\num{255} \pm \num{28}$ & $\num{0.52} \pm \num{0.06}$ & \num{26.9} & \num{23} \\
    L4--A & $\num{30.8} \pm \num{2.0}$ & $\num{0.25} \pm \num{0.02}$ & \num{20.9} & \num{25} \\
    L4--B & $\num{19.7} \pm \num{1.5}$ & $\num{0.17} \pm \num{0.01}$ & \num{20.6} & \num{23} \\
    \bottomrule
  \end{tabular}
  \label{tab:background}
\end{table}

The first measurement we discuss is a sampling of the background rate, at a frequency of approximately \SI{1}{\hertz} and for a total time of \SI{24}{\hour} for each of the links under investigation.
The reported values account for the known efficiency and dark count rate of the detectors, so that they represent the actual background photon rate in the fiber.
The wavelength band of this unfiltered acquisition is determined by the spectral response of the SNSPDs, which is several hundred nanometer wide around the design wavelength of \SI{1550}{\nano\meter}.
Since the detection efficiency is not constant over this range, we calculate the effective efficiency by averaging the efficiency curve provided by the manufacturer, weighted by the measured noise spectrum, which we discuss later.
Given that our characterization covers only partially the wavelength range of the light and the detector efficiency curve, we expect a residual inaccuracy in the value of the efficiency, which is reflected in our estimation of the absolute background rate.
However, this does not affect our most relevant conclusions.

\begin{figure}
  \centering
  \begin{subfigure}[t]{0.48\textwidth}
    \centering
    \includegraphics[width=\textwidth]{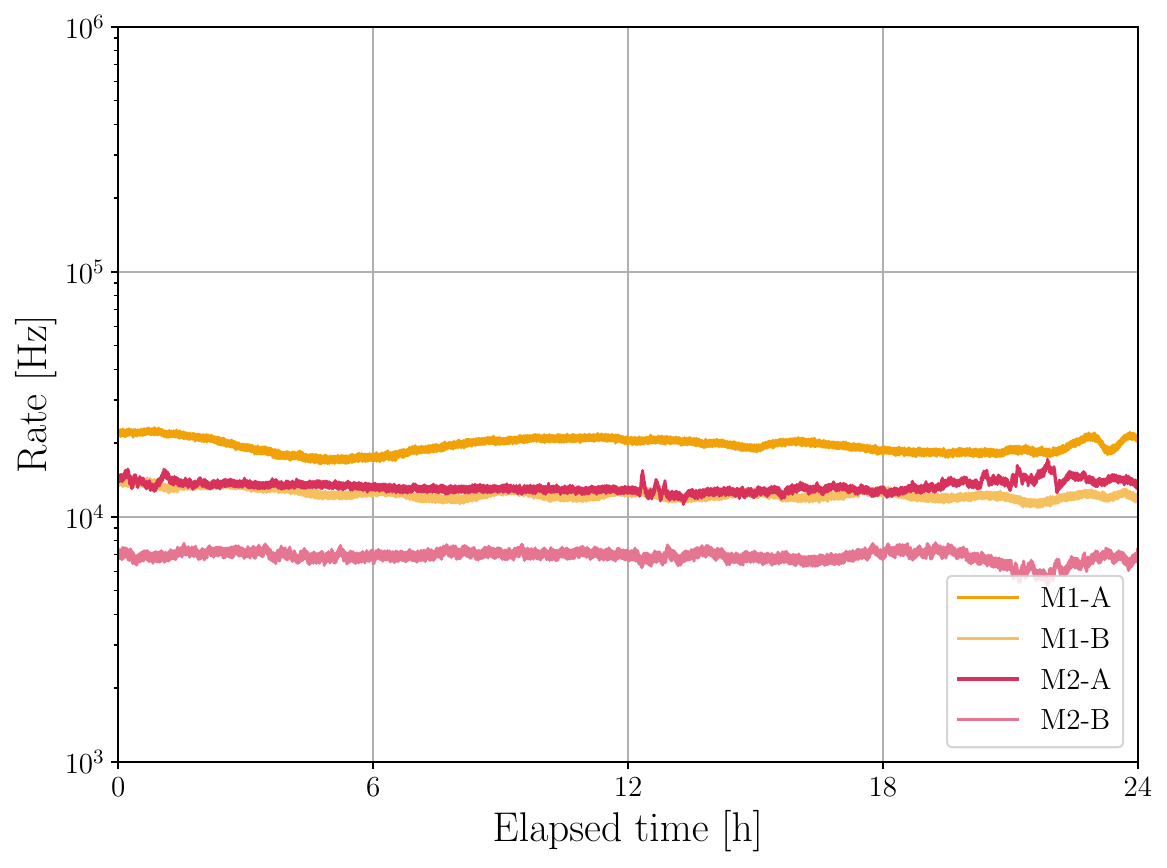}
    \caption{Metropolitan links.}
    \label{fig:bkg_long_rate_1}
  \end{subfigure}
  \hfill
  \begin{subfigure}[t]{0.48\textwidth}
    \centering
    \includegraphics[width=\textwidth]{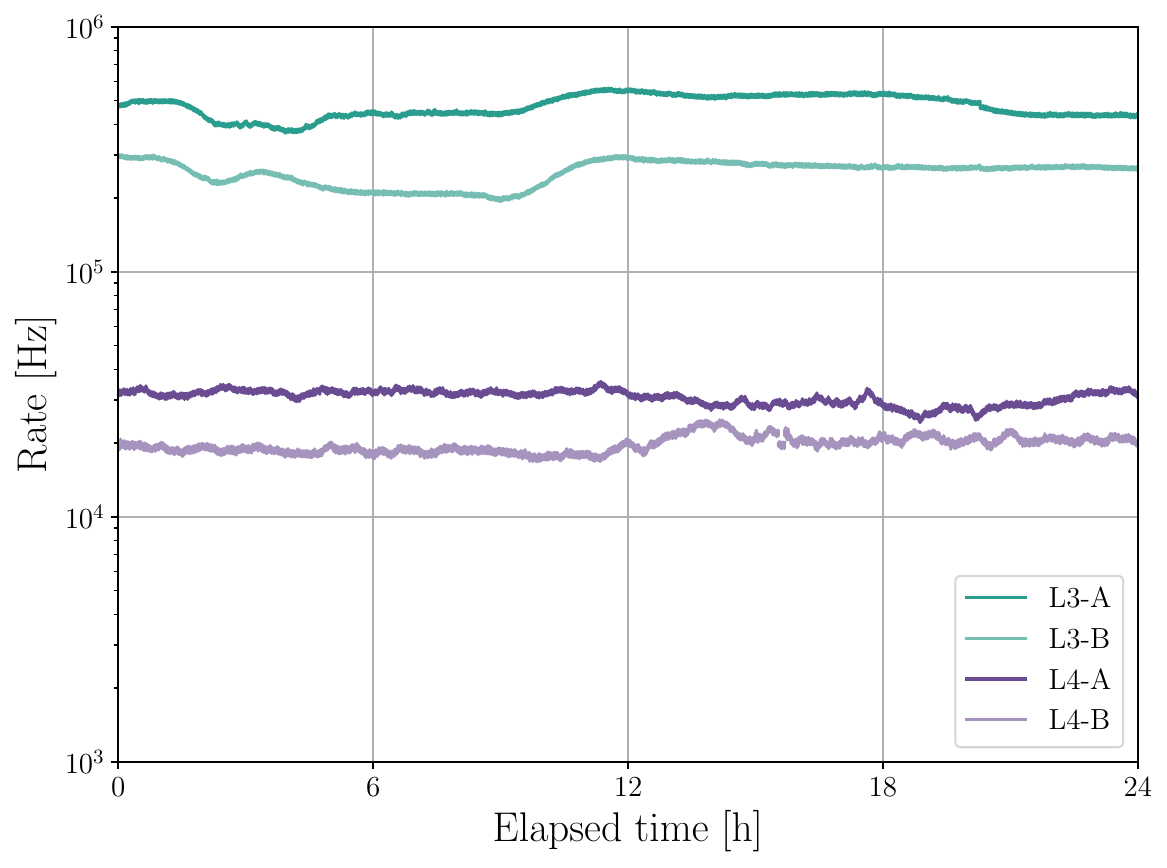}
    \caption{Long-distance links.}
    \label{fig:bkg_long_rate_2}
  \end{subfigure}
  \caption{Background photon rate measured in the deployed fiber links of the NQCIS network. The two strands in each link are measured simultaneously. Duration: \SI{24}{\hour} for each link; sampling rate: $\sim\SI{1}{\hertz}$.}
  \label{fig:bkg_long_rate}
\end{figure}

Figure \ref{fig:bkg_long_rate} shows the result of the full sampling, while averages and standard deviations are reported in Tab.~\ref{tab:background}.
The measured rates differ by orders of magnitude from link to link, ranging from \SI{6.9 \pm 0.3}{\kilo\hertz} on M2-B to \SI{476 \pm 49}{\kilo\hertz} on L3-A.
Furthermore, they do not scale regularly with length, since they are slightly larger in L4 compared to the metropolitan links, but much lower than in the shorter L3 link.
This suggests that the main source of noise is not something natural, like black-body radiation, but rather artificial, such as optical communication in neighboring fibers, which depends on the usage conditions of these strands by other users.

Then, we insert a band-pass filter matched with channel 34 of the ITU grid, which is centered at \SI{1550.12}{\nano\meter} and has a bandwidth of \SI{0.8}{\nano\meter} (Lightcomm Technology DWDM-1x2-100G-CH34).
This wavelength choice is arbitrary, dictated by the particular QKD devices deployed in NQCIS, but it exemplifies the general effectiveness of such filters.
We include in Tab.~\ref{tab:background} the average and standard deviation of the measured samples, and the extinction ratio (in \unit{\dB}) with respect to the unfiltered rate.
We can see that a single filter can reduce the background rate by more than three orders of magnitude in the M2 link, but loses effectiveness in other links, particularly M1, where it reaches only about \SI{15}{\dB}.

The reason for this variability becomes clearer by observing the wavelength spectrum of the noise, which we measured by inserting a tunable band-pass wavelength filter (Agiltron FOTF) in the setup of Fig.~\ref{fig:background_noise_setup}.
Because this filter is only guaranteed to work in the \SIrange{1510}{1660}{\nano\meter} range, and in particular is transmissive at \SI{1310}{\nano\meter}, we expect our values to be affected by a residual additive term.
However, we can still evaluate the relative probability density of the noise over a wide wavelength range that covers much more than the optical C-band.

\begin{figure}
  \centering
  \begin{subfigure}[t]{0.48\textwidth}
    \centering
    \includegraphics[width=\textwidth]{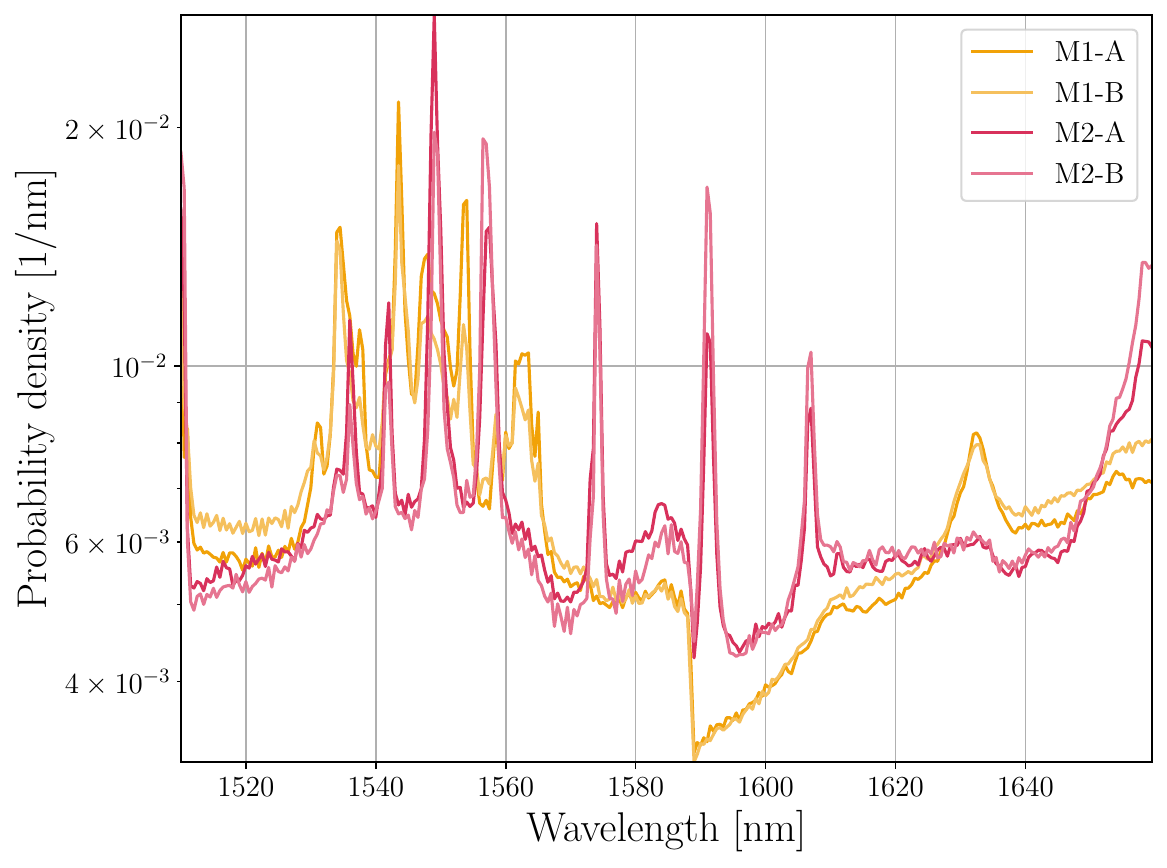}
    \caption{Metropolitan links.}
    \label{fig:bkg_filter_prob_density_1}
  \end{subfigure}
  \hfill
  \begin{subfigure}[t]{0.48\textwidth}
    \centering
    \includegraphics[width=\textwidth]{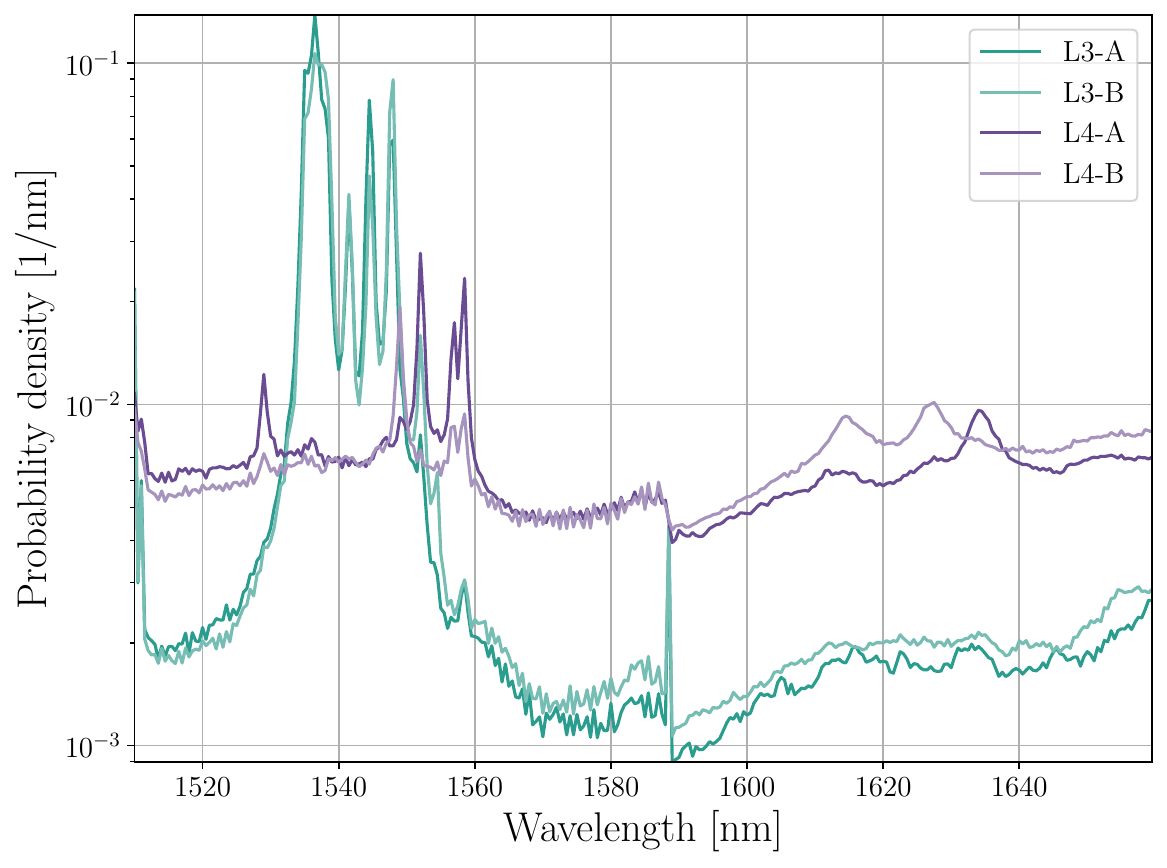}
    \caption{Long-distance links.}
    \label{fig:bkg_filter_prob_density_2}
  \end{subfigure}
  \caption{Relative probability density of the background photon rate over wavelength in the deployed fiber links of the NQCIS network.}
  \label{fig:bkg_filter_prob_density}
\end{figure}

The result, shown in Fig.~\ref{fig:bkg_filter_prob_density}, features clear peaks, which are indicative of specific wavelength channels being in use in nearby strands.
The distribution of the peaks changes from link to link, which is compatible with the fact that optical communication channels are generally allocated arbitrarily, so that different ones are used in different links.
For instance, the M1 link shows prominent peaks at \num{1535}, \num{1543}, and \SI{1553}{\nano\meter}, the M2 link at \num{1549}, \num{1574}, and \SI{1591}{\nano\meter}, the L3 link shows a series of five peaks between \num{1535} and \SI{1552}{\nano\meter}, and the L4 link three peaks between \num{1552} and \SI{1559}{\nano\meter}, but also broader, weaker peaks between \num{1614} and \SI{1634}{\nano\meter}.

There is a good correspondence between the peaks in the spectra of the two strands over the same link, which is consistent with the fact that they are colocated in the same cable and hence subject to similar evanescent coupling from neighboring strands.
However, this is not completely true for the L4 link, where peaks at \num{1552} and \SI{1634}{\nano\meter} appear only in the A strand, whereas peaks at \num{1614} and \SI{1627}{\nano\meter} are present only in the B strand.
We note that the L4 link is rented from a different provider with respect to the other three, which might indicate a different cabling structure, resulting in a different pattern of light leakage.

Even on the same strand, the background rate can differ by more than one order of magnitude from wavelength to wavelength, even for almost adjacent channels.
For instance, the L3-B link reports a rate \num{10.5} times higher at \SI{1548}{\nano\meter} compared to \SI{1550}{\nano\meter}.
This indicates that the effectiveness of spectral filtering can vary significantly depending on the specific wavelength channel used for the quantum transmission.
Therefore, when deploying QKD in a fiber link, it can be beneficial to know or observe the usage pattern of neighboring strands, to select a wavelength channel that is less affected.
This can be more important than the modest gains that can be obtained by choosing the wavelength which minimizes attenuation (Fig.~\ref{fig:att_vs_wl}), and can significantly move $Y_0/\eta$ away from the boundary for $\SKR=\num{0}$ (Fig.~\ref{fig:skr_contour_vs_y0_q}).
Unfortunately, this also means that changes in the usage of other strands can affect QKD performance even after deployment.

We note that the extinction ratios of Tab.~\ref{tab:background} do not align with the probability distributions in Fig.~\ref{fig:bkg_filter_prob_density}.
For instance, for the M2-A link, the extinction ratio is above \SI{30}{\dB}, whereas the probability density is greater than \SI{0.01}{\per\nano\meter} around \SI{1550}{\nano\meter}.
We attribute this to the fact that the latter is normalized to its integral in a restricted wavelength range, while the former is related to the total rate over the full detection range of the SNSPDs.
This is a good indication of the presence of background light outside the range of the tunable filter, and suggests eliminating more wavelength bands with additional filters to improve the accuracy of the characterization.

Finally, we investigate the degree of polarization (DOP) of the background light by inserting an additional MPC in the dashed box of Fig.~\ref{fig:background_noise_setup} and adjusting it to find the maximum extinction ratio between the two detectors.
If we assume that the PBS is operating an ideal projective measurement of polarization, the maximum extinction rate $r_{ext}$ is related to the DOP of the incoming light by the relation $\DOP = (r_{ext}-1)/(r_{ext}+1)$.

Table~\ref{tab:background} reports the results of this measurement for each fiber.
The DOPs are all under \SI{50}{\percent}, which is compatible with the fact that the background light is generated by multiple sources, likely to be multiple communication channels in nearby fibers, and therefore is a mixture of different polarization states.
Two strands in the same link usually have similar DOPs, in agreement with the previous conclusion, backed by the spectra of Fig.~\ref{fig:bkg_filter_prob_density}, that they are affected by similar sources of background light.
We note that these values are obtained with measurements in quick succession, likely to approximate the instantaneous DOP of the background light.
However, observation of the subdivision of the background rate across the two exit arms of the PBS over time indicates that the state of polarization exhibits slow drifts.
This means that the effective DOP averaged over longer timescales is likely to be lower than the values we report.

\subsection{Polarization effects}
\label{sec:polarization}

Because of birefringence, the fiber alters the state of polarization (SOP) of light traveling through it.
We focus on two distinct physical phenomena: drift and polarization-mode dispersion (PMD).
Drift refers to the movement of the SOP over time, which is due to environmental factors such as temperature fluctuations or mechanical vibrations \cite{Imai1987}.
This leads to a time-varying misalignment between the prepared and measured polarization states, which must be compensated with active control systems.
PMD is the result of the fact that different polarization modes propagate at different speeds in the fiber, leading to a temporal broadening of light pulses and a loss of purity of the SOP \cite{Gordon2000,Galtarossa2005}.
Countermeasures against PMD also exist and are usually based on strongly birefringent segments that are inserted before or after a link to compensate for the differential group delay (DGD) between the two polarization modes \cite{Sunnerud2000}.

The two phenomena work at very different timescales.
As we will see, drift operates over minutes or hours in the links described here (Sec.~\ref{sec:drift}) and over fractions of a second to days in the literature \cite{Ding2017,Czegledi2016}.
In any case, drift is several orders of magnitudes slower than the characteristic durations of QKD pulses, which are typically in the order of \SI{100}{\pico\second} to \SI{1}{\nano\second}, reaching down to the tens of picoseconds in extreme cases \cite{Woodward2021,Morrison2023}.
Therefore, its action on each quantum state can be very well approximated by a unitary transformation.
Only when considering an ensemble of pulses transmitted over a long time, drift causes an effective depolarization.
Hence, the relevant timescales to compare drift against are those characteristic of the accumulation of signals to create a key, or of the compensation system deployed to counter the drift itself.
Instead, the characteristic time of PMD is the DGD, which is limited to a few picoseconds even for very long fibers (Sec.~\ref{sec:pmd}).
Because this is small compared to QKD pulses, the depolarizing effect remains usually inconsequential.
This might change in the future, if QKD devices are developed to use shorter pulses, but the practicality of this might still be limited by spectral effects and inadequate detector resolution.

Nonetheless, characterizing these effects is important, so as to predict the performance of deployments and to put quantitative requirements on drift compensating systems and QKD pulses.
This is not only true for polarization-encoded QKD protocols, but also for any scheme that relies on interference between two traveling quantum signals or one signal and a local oscillator, such as twin-field QKD or continuous-variable QKD.
In such single-polarization architectures, the receiver can use a polarizer to convert polarization noise into an effective loss, but in any case the performance is affected.

\subsubsection{Polarization drift}
\label{sec:drift}

The previous considerations on timescales allow studying drift at slow rates compared to the characteristic time of a QKD pulse, but fast enough to be comparable with those of compensation systems, typically in the order of \SI{1}{\hertz} \cite{Chen2007}.
Consequently, we use a continuous-wave laser (IDPhotonics Cobrite DX2-LC) at the fixed wavelength of \SI{1550}{\nano\meter} and measure the SOP at the output of the link with a polarimeter (Thorlabs PAX1000IR2), at a sampling period of $\delta t\approx \SI{0.1}{\second}$.
The polarimeter reports the four Stokes parameters $S_0, S_1, S_2, S_3$ and we calculate the Bloch vector of the quantum state as $\vec{n} = (n_1, n_2, n_3) = (S_1, S_2, S_3)/S_0$.
The state in the Hilbert space of polarization is then given by the density matrix $\varrho = \frac{1}{2}(\mathbb{I} + \vec{n}\cdot\vec{\sigma})$, where $\vec{\sigma}$ is the vector of Pauli matrices.

We report a full analysis for two links, M2 and L3, in a looped configuration (length about \SI{39}{\km} and \SI{162}{\km}, respectively), for which we collected one week of data.
We observe that the DOP, which coincides with $|\vec{n}|$, is compatible with \num{1} within the precision of the polarimeter, and similar to the one found by analyzing directly the output of the laser, without any link.
This validates the hypothesis that the drift is unitary at the timescales of the measurement (\SI{0.1}{\second}), and even more so at those of a QKD pulse, which is confirmed also by the spectral profile of the drift, discussed in what follows.
Therefore, we approximate $\vec{n}\approx\hat{n}$ (with the hat indicating unit norm) and focus on the motion on the surface of the Bloch sphere, without any radial component.

\begin{figure}
  \centering
  \begin{subfigure}[t]{0.48\textwidth}
    \centering
    \includegraphics[width=\textwidth]{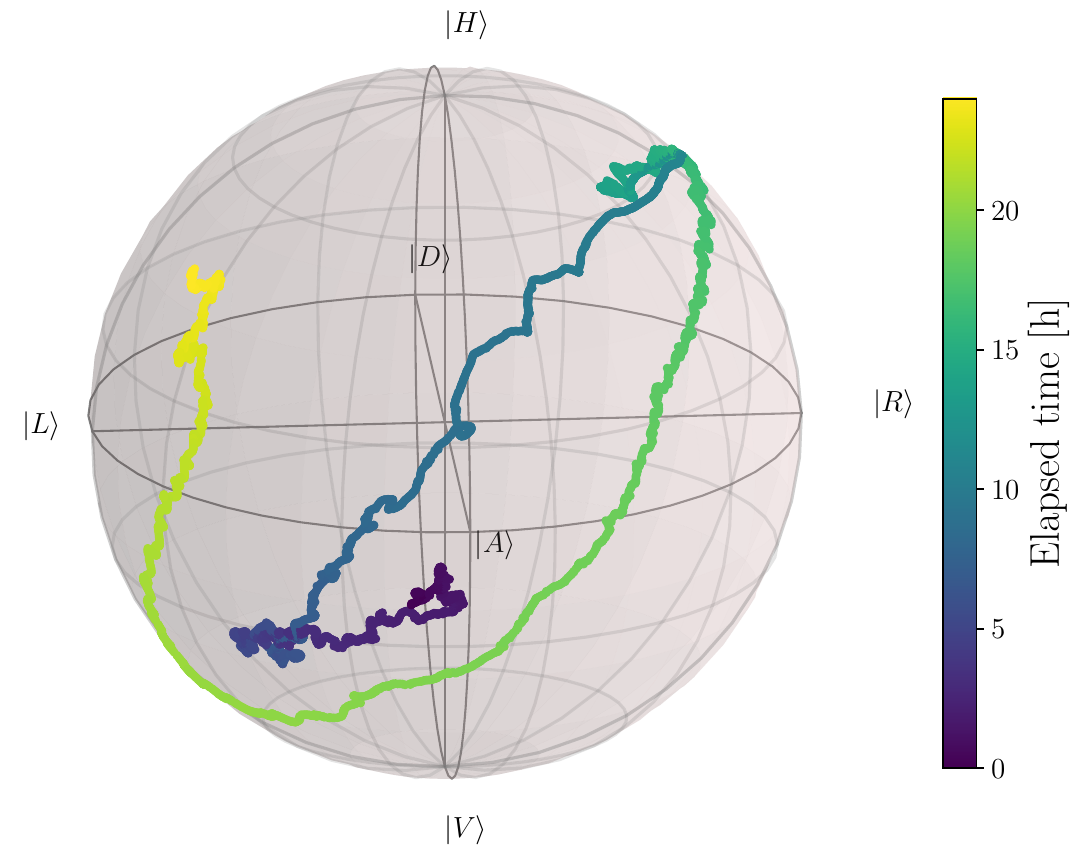}
    \caption{M2 link.}
    \label{fig:polar_plot_bloch_ris_loop_long_pure_t24h_d10s}
  \end{subfigure}
  \hfill
  \begin{subfigure}[t]{0.48\textwidth}
    \centering
    \includegraphics[width=\textwidth]{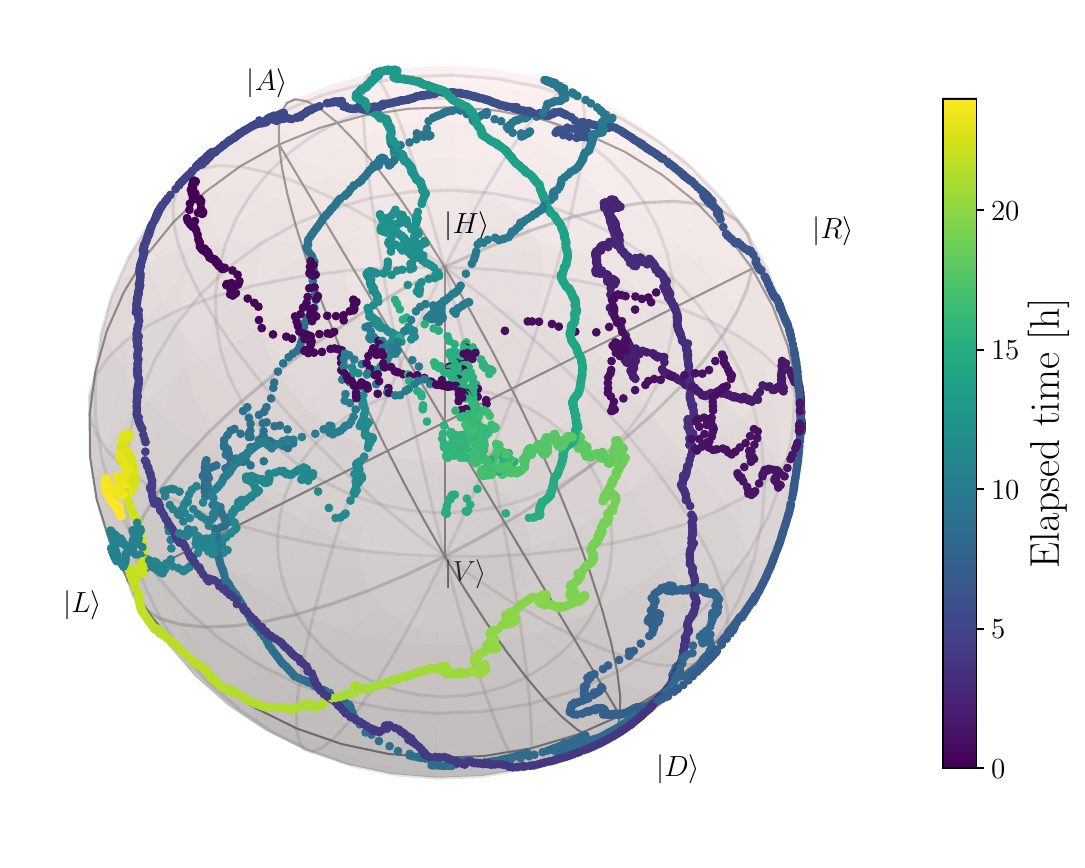}
    \caption{L3 link.}
    \label{fig:polar_plot_bloch_nor_loop_long_pure_t24h_d10s}
  \end{subfigure}
  \caption{Evolution of the Bloch vector over \SI{24}{\hour}.}
  \label{fig:polar_plot_bloch}
\end{figure}

Figure~\ref{fig:polar_plot_bloch} shows the evolution of the Bloch unit vector over the first \SI{24}{\hour} of the measurement, highlighting a pronounced drift away from the initial point.
The two trajectories are quantitatively different, with the drift being wider, faster, and more irregular in the L3 link.
This stresses the importance of two points: first of all, a compensation system is necessary to counter this drift, otherwise the polarization would get away from the desired state in less than one hour, and secondly, the quantitative properties of the drift are link-dependent and must be characterized individually.

\begin{figure}
  \centering
  \includegraphics[width=0.5\textwidth]{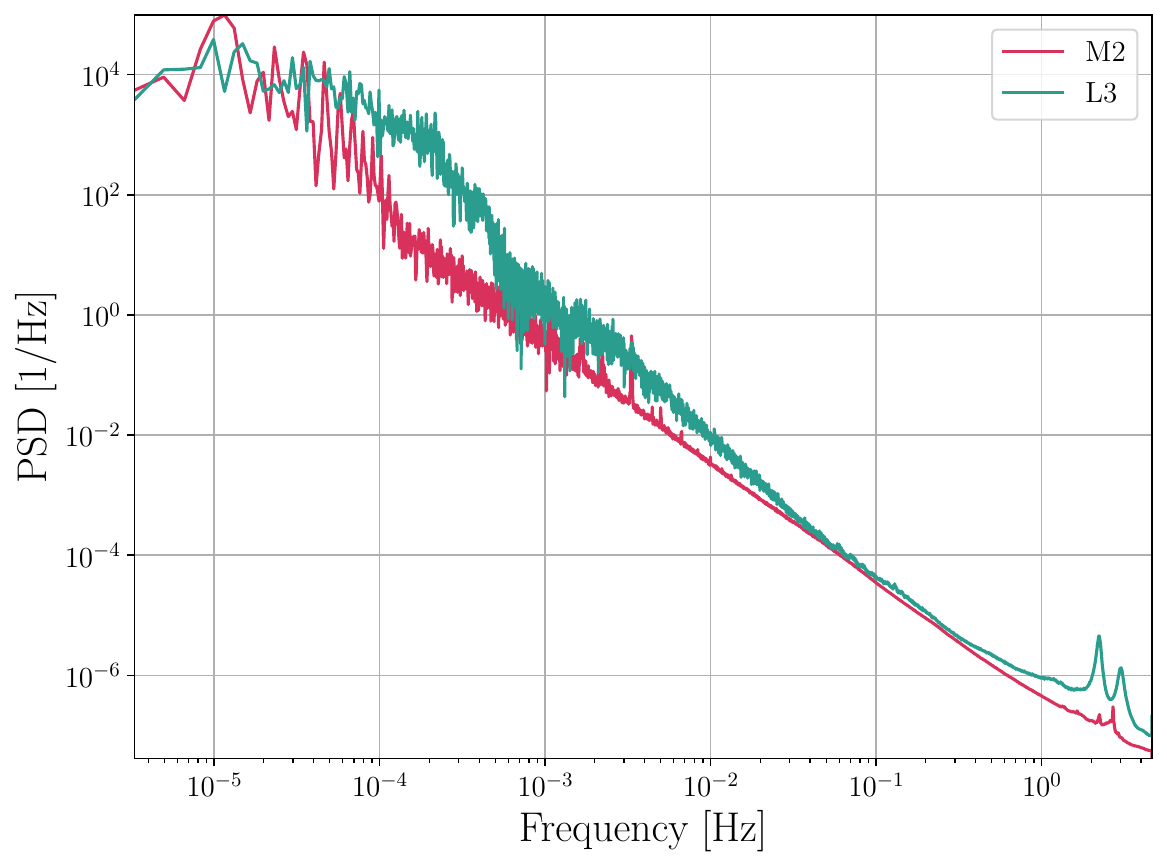}
  \caption{Power spectral density of the Bloch vector components for the M2 and L3 links.
  To ease visualization, after calculating the PSD, the frequency axis is binned with logarithmic spacing (\num{1000} points per decade) and values corresponding to frequencies within each bin are averaged.}
    \label{fig:polar_plot_multi_periodogram}
\end{figure}

Previous literature has described the SOP drift as a random walk, following a model of the fiber as a sequence of randomly oriented and independent birefringent segments \cite{Imai1987,Imai1988,Karlsson2000,Czegledi2016,Czegledi2017}.
In what follows, we report how our data confirms that this is a valid approximation but also shows some deviations from random-walk behavior, which may be relevant for the design of compensation systems.

Figure~\ref{fig:polar_plot_multi_periodogram} displays the sum of the power spectral densities (PSD) of each Bloch unit vector component, for a full week of continuous measurement.
The low-frequency plateau is due to the constrained geometry of the sphere and is a property of any trajectory that is protracted over a time long enough to explore the whole surface.
The subsequent roll-off shows a $1/f^2$ trend (as evidenced by the slope of about $\num{-2}$ in the log-log plot), which is compatible with a random walk.
Although our Nyquist frequency is limited by the sampling rate, the descending trend indicates that drifts at higher frequencies are likely to be inconsequential, validating the unitarity assumptions at the small timescales of QKD pulses.

In addition to the $1/f^2$ behavior, there are also some peaks at specific frequencies, which indicate the presence of periodic components in the drift.
In particular, the L3 link displays peaks at approximately \SI{2}{\hertz} and \SI{3}{\hertz}, which are much less pronounced in the M2 link.
This difference is a good indication that these peaks are not induced by our measurement setup, but should rather be attributed to mechanical vibrations acting on the fiber, which are different from link to link.

\begin{figure}
  \centering
  \begin{subfigure}[t]{0.48\textwidth}
    \centering
    \includegraphics[width=\textwidth]{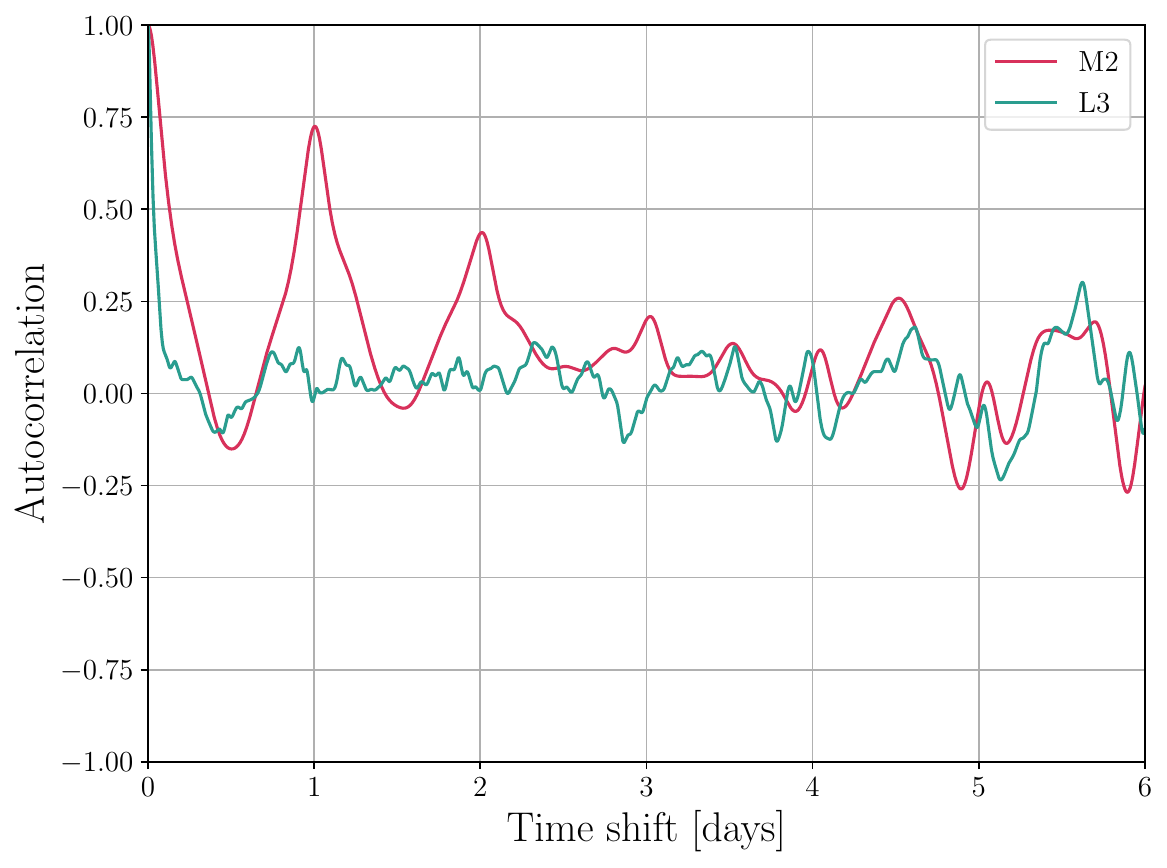}
    \caption{First six days.}
    \label{fig:polar_plot_multi_autocorr_long}
  \end{subfigure}
  \hfill
  \begin{subfigure}[t]{0.48\textwidth}
    \centering
    \includegraphics[width=\textwidth]{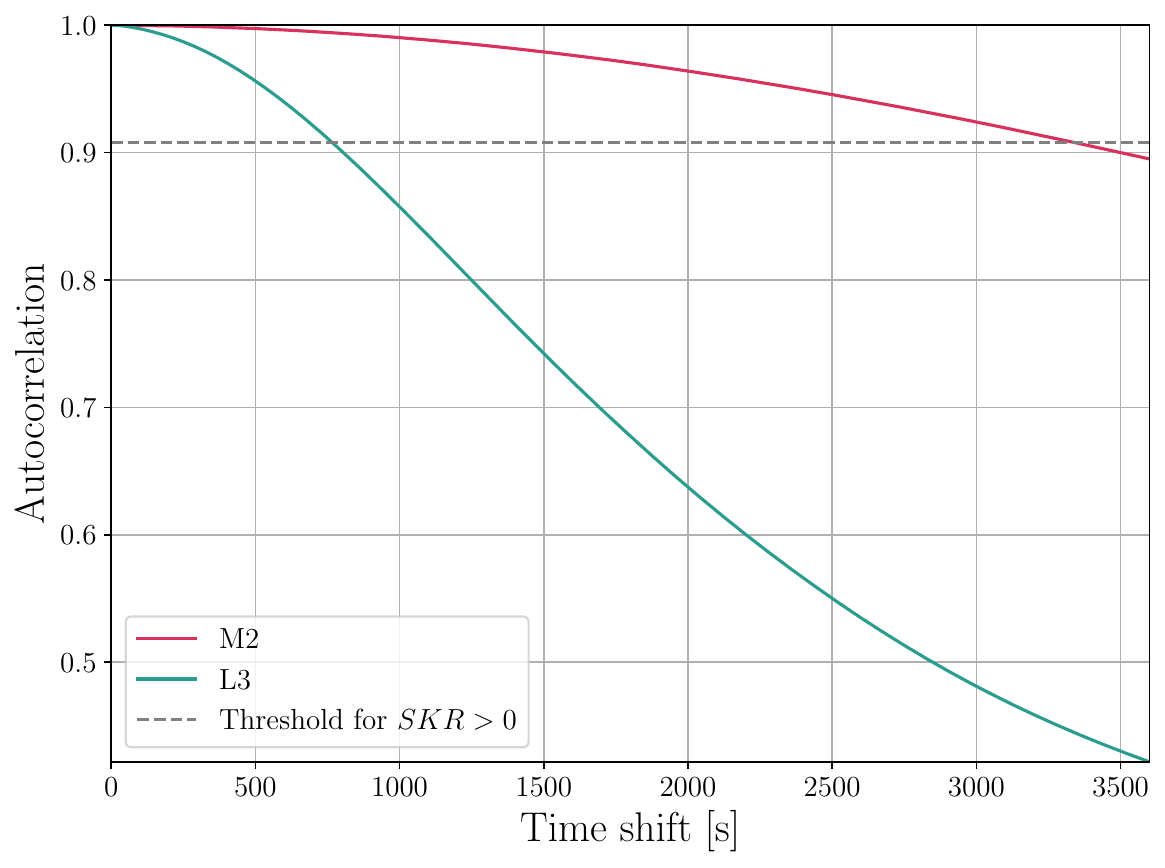}
    \caption{First hour.}
    \label{fig:polar_plot_multi_autocorr_short}
  \end{subfigure}
  \caption{Autocorrelation of the Bloch vector components for the M2 and L3 links for a one-week-long experiment at different time scales.}
\end{figure}

The M2 link shows a peak at approximately $\SI{1.1e-5}{\hertz} \approx 1/\SI{24}{\hour}$, which is compatible with the daily temperature cycle.
This is even more evident by observing the autocorrelations of the components and their sum, calculated from the discrete-time measurements as follows:
\begin{equation}
  \begin{aligned}
    A_{i}(l\delta t) &= \frac{1}{N-l} \sum_{k=0}^{N-l-1} \hat{n}_i(k\delta t) \cdot \hat{n}_i((k+l)\delta t), \\
    A(l\delta t) &= \sum_{i=1}^{3} A_i(l\delta t) = \frac{1}{N-l} \sum_{k=0}^{N-l-1} \hat{n}(k\delta t) \cdot \hat{n}((k+l)\delta t),
  \end{aligned}
\end{equation}
where $N$ is the total number of samples, $\delta t$ is the sampling period (approximately \SI{0.1}{\second}) and $\cdot$ is the Euclidean inner product.
These autocorrelations can be straightforwardly calculated numerically as the inverse discrete Fourier transforms of the PSD of each polarization component, up to a normalization factor.
Figure~\ref{fig:polar_plot_multi_autocorr_long}, plots $A(t)$ for delays up to six days (for more than this, the estimation loses precision due to the small number of available data points).
The autocorrelation of the M2 link shows a clear periodicity of \SI{24}{\hour}, with the state of polarization returning closer to itself after one day.
This is less pronounced in the L3 link, where the autocorrelation decays more rapidly.
This difference might be attributed to different exposure to the environment in each deployed link.

Another relevant deviation from a pure random-walk behavior is shown in Fig.~\ref{fig:polar_plot_multi_autocorr_short}, which zooms in on the autocorrelation curves, for delays of up to \SI{1}{\hour}.
The initial decay is clearly parabolic, whereas random-walk theory would predict an exponential drop, which is linear to first order for short times.
This is an indication of the fact that consecutive steps of the walk maintain some correlation, consistent with a driving force caused by environmental factors that change very slowly over time.
This introduces some regularity in the trajectory, qualitatively appreciable especially in Fig.~\ref{fig:polar_plot_bloch_ris_loop_long_pure_t24h_d10s}.

\begin{figure}
  \centering
  \includegraphics[width=0.5\textwidth]{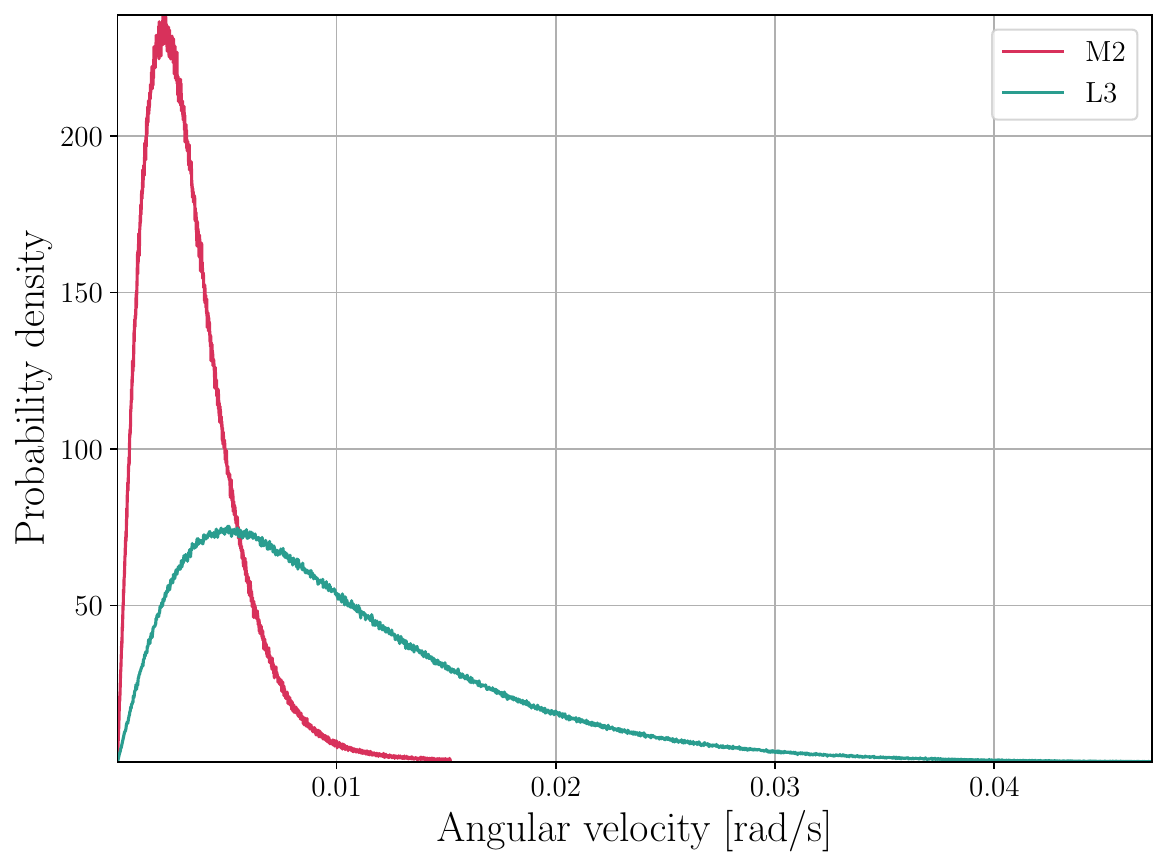}
  \caption{Probability distribution of the angular velocity of the Bloch vector for the M2 and L3 links.}
  \label{fig:polar_plot_multi_angular_speed_distribution}
\end{figure}

Finally, Fig.~\ref{fig:polar_plot_multi_angular_speed_distribution} shows the distribution of angular velocity for each link.
In standard random-walk theory with independent, normally-distributed increments, this should follow a Rayleigh distribution.
While the experimental data exhibits the characteristic linear scaling of the Rayleigh distribution at small velocities, it displays a higher peak and a slower-decaying tail.
This behavior is better approximated by a Weibull distribution with shape parameter $k \approx 1.6$ (where $k = 2$ corresponds to the Rayleigh case).
This is consistent with a drift caused by several concurrent factors, with the increments being locally normally distributed, but governed by a fluctuating variance.
Figure~\ref{fig:polar_plot_multi_angular_speed_distribution} also confirms quantitatively that the M2 link drifts more slowly than the L3 link, which could only qualitatively be inferred from Fig.~\ref{fig:polar_plot_bloch}.
Indeed, the average angular velocity is about \SI{3}{\milli\radian\per\second} for the M2 link, and \SI{10}{\milli\radian\per\second} for the L3 link.

We now want to relate these results with a meaningful estimation of $q$.
Of course, this is highly dependent on the control mechanisms that are in place to counter the drift, but our analysis of the free-running phenomenon can give a good idea of the requirements for such mechanisms.
We first note that if we model the measurement operated by the receiver as a projective measurement which is aligned with the SOP Bloch vector at $t=0$, the expected value of the QBER after a delay $\tau$ during which the SOP drifts is related to $A(\tau)$ as follows:
\begin{equation}
  \langle q_{\mathrm{drift}}(\tau) \rangle = \frac{1}{2} \left( 1 - \langle \hat{n}(t) \cdot \hat{n}(t+\tau) \rangle \right) = \frac12\left( 1 - A(\tau) \right),
  \label{eq:qdrift}
\end{equation}
where the expectation value is taken over the time $t$.

Previous literature has suggested using the time needed to reach meaningful QBER thresholds as a figure of merit for the drift \cite{Ding2017}.
In our case, we can see that the M2 link reaches the previously defined threshold of \SI{4.6}{\percent} after approximately \SI{3300}{\second}, while the L3 link takes \SI{750}{\second} (see the dotted line in Fig.~\ref{fig:polar_plot_multi_autocorr_short}).
These durations indicate the mean time at which QKD would stop operating in the absence of any compensation system.
However, if we include such a system and consider that its characteristic operation time is much smaller, we rather suggest using the latter as $\tau$ in Eq.~\eqref{eq:qdrift} to estimate the residual $q$ after compensation.
For example, considering a worst-case scenario of a qubit-based compensation system that must integrate detections for $\tau\approx\SI{1}{\minute}$ to have a meaningful signal-to-noise ratio, we find $\langle q_{\mathrm{drift}}(\tau) \rangle = \num{4e-5}$ and $\langle q_{\mathrm{drift}}(\tau) \rangle = \num{6e-4}$ for the M2 and L3 links, respectively.
These very small values are indicative of the relatively low requirements that the link poses on drift-compensation systems, but should be considered a lower bound, destined to increase when a more realistic model of the compensation system is taken into account.

\subsubsection{Polarization-mode dispersion}
\label{sec:pmd}

\begin{figure}
  \centering
  \includegraphics[width=0.5\textwidth]{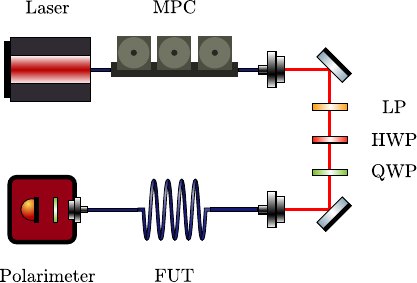}
  \caption{Scheme of the measurement setup for the PMD measurement in deployed fiber links.
  MPC: manual polarization controller, LP: linear polarizer, HWP: half-wave plate, QWP: quarter-wave plate, FUT: fiber link under test.}
  \label{fig:pmd_setup}
\end{figure}

The characterization of PMD is quite common in the telecom industry because it can limit the bandwidth of optical communication systems.
In the field of quantum optics, a popular technique to study it is Jones matrix eigenvector analysis, which aims at reconstructing the transformation matrix of the fiber link at different wavelengths and extract the DGD from it \cite{Poole1986,Heffner1992,Gordon2000}.
The setup we employ for this is shown in Fig.~\ref{fig:pmd_setup}.
A continuous-wave wavelength-tunable laser source (Yenista Optics Tunics T100S) is used to prepare the input light.
A free-space assembly of a linear polarizer (Thorlabs LPNIR050-MP2), a half-wave plate (Thorlabs WPHSM05-1550), and a  quarter-wave plate (Thorlabs WPQSM05-1550) is used to switch the polarization to three different states, i.e., horizontal, vertical, and diagonal (defined up to a global unitary transformation).
Light prepared in this way is sent through the link of interest and the output SOP is measured with the same polarimeter as before.
Since the waveplates are mounted on motorized rotation stages (Thorlabs DDR25/M), we switch automatically between the three input states and scan the wavelength of the laser source in a range from \num{1540} to \SI{1560}{\nano\meter}, so as to obtain the Jones transformation matrix as a function of wavelength.
This scan, which takes approximately \SI{4}{\minute}, is repeated for \SI{48}{\hour}.

\begin{figure}
  \centering
  \begin{subfigure}[t]{0.48\textwidth}
    \centering
    \includegraphics[width=\textwidth]{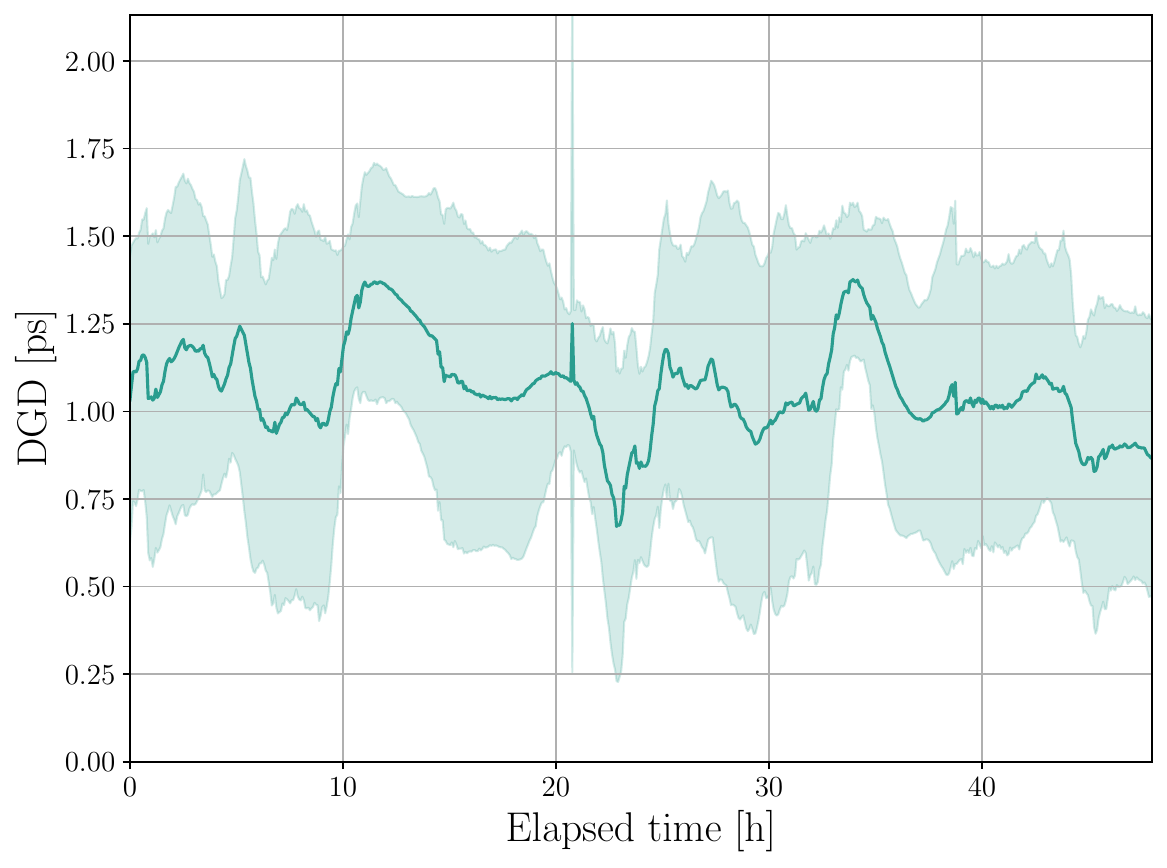}
    \caption{Evolution over time, after wavelength averaging.}
    \label{fig:pmd_plot_dgd_vs_time}
  \end{subfigure}
  \hfill
  \begin{subfigure}[t]{0.48\textwidth}
    \centering
    \includegraphics[width=\textwidth]{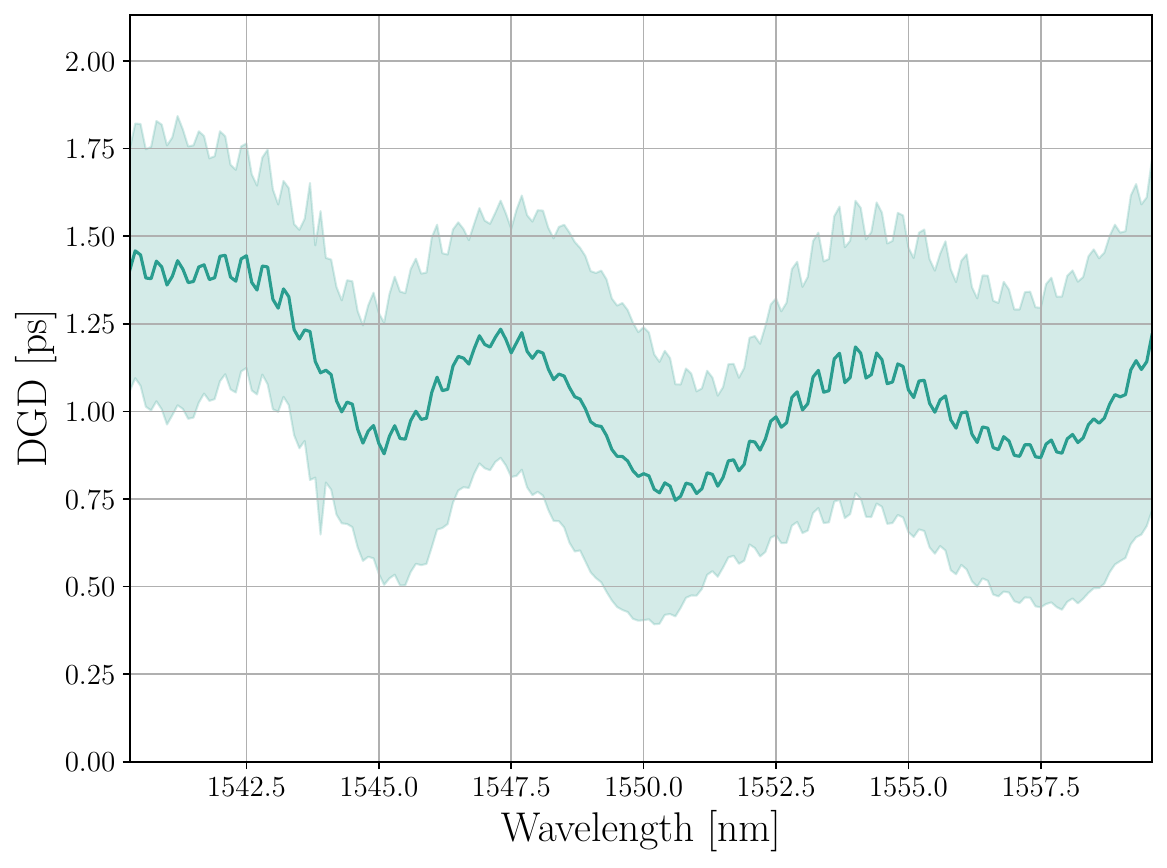}
    \caption{Evolution against wavelength, after time averaging.}
    \label{fig:pmd_plot_dgd_vs_wavelength}
  \end{subfigure}
  \caption{DGD for the L3 link, looped at the remote end.}
\end{figure}

Following the algorithm described in Ref.~\cite{Gordon2000}, we use the data to find the PMD vector, whose direction indicates the principal states of polarization (those states for which the action of the transformation matrix is independent of the wavelength at first order) and whose modulus is the DGD.
All is expressed as a function of both wavelength and time.
Figure~\ref{fig:pmd_plot_dgd_vs_time} shows the evolution of the DGD over time, after wavelength averaging, whereas Fig.~\ref{fig:pmd_plot_dgd_vs_wavelength} shows the DGD versus wavelength, after time averaging.
Both graphs refer to the L3 link, looped at the remote end.
We can see that the DGD shows a slow and noisy oscillation, without sudden or wide changes (except for one temporary jump after approximately \SI{20}{\hour}).
For the particular wavelength of \SI{1550}{\nano\meter}, the DGD shows a mean of \SI{0.81}{\pico\second} and a standard deviation of \SI{0.41}{\pico\second}.

According to the aforementioned model for which the fiber is composed of several birefringent segments, the DGD scales with the square root of the length.
Consequently, it is common to express it in terms of a coefficient in \unit{\pico\second\per\kilo\meter^{1/2}}, obtained by dividing it by the square root of the link length, yielding \SI{0.06}{\pico\second\per\kilo\meter^{1/2}} for this particular case.
This value is less than the maximum specified by the ITU standard for modern optical fibres, which is \SI{0.2}{\pico\second\per\kilo\meter^{1/2}}.
For the one-way unlooped link, we can expect a further reduction of the DGD by a factor of $\sqrt{2}$.

To quantify the effect of PMD on $q$, we need to consider that the relevant physical mechanism is the entanglement of the polarization state with the time-of-arrival degree of freedom.
The amount of entanglement depends on the temporal profile of the signal pulse, and on the alignment between the initial polarization state and the principal states of polarization \cite{Antonelli2011}.
Assuming that such alignment is uncontrolled, the expected shortening of the Bloch vector due to PMD is equal to the autocorrelation of the time amplitude distribution, evaluated at the DGD.
Therefore, considering for instance a Gaussian-shaped time intensity distribution of standard deviation $\sigma_t$ and a DGD of $\tau$, the expected QBER is:
\begin{equation}
  \langle q_{\mathrm{PMD}} \rangle = \frac12 \left(1-e^{-\frac{\tau^2}{8\sigma_t^2}}\right).
  \label{eq:qpmd}
\end{equation}
For example, for $\sigma_t=\SI{100}{\pico\second}$, the measured value $\tau=\SI{0.81}{\pico\second}$ yields $\langle q_{\mathrm{PMD}} \rangle \approx \num{4e-6}$.
Again, this low value is indicative of the small effect that PMD has on current polarization-encoded QKD, but potential technology developments in terms of faster pulses (smaller $\sigma_t$) would make the effect more significant in the QKD context.

\subsubsection{Combining the contributions}
\label{sec:combining}

Even when combined assuming simplistically independent depolarizing channels, the link-induced QBER on the L3 link is still very small:
\begin{equation}
  q_{\mathrm{link}} = q_{\mathrm{drift}} + q_{\mathrm{PMD}} - 2 q_{\mathrm{drift}} q_{\mathrm{PMD}} \lesssim \num{1e-3}\,.
  \label{eq:qlink}
\end{equation}
In these conditions, it is likely for the final value of $q$, Eq.~\eqref{eq:q}, to be dominated by the imperfections of the QKD devices, arising from non-ideal state preparation and non-projective measurements:
\begin{equation}
  q_{\mathrm{syst}} = q_{\mathrm{prep}} + q_{\mathrm{meas}} - 2 q_{\mathrm{prep}} q_{\mathrm{meas}}\,.
  \label{eq:qsyst}
\end{equation}
Manufacturers can determine $q_{\mathrm{syst}}$ by measuring $Q_\mu$ using a trivial channel and moving backwards in Fig.~\ref{fig:graph_analysis} to eliminate the terms due to $Y_0$ and $\eta$.
Even if these values are not normally disclosed to the users, manufacturers could employ them internally with user-provided $q_{\mathrm{link}}$ to assess the ultimate performance.
However, we must stress that our simplified model does not account for any interplay between the different contributions to $q$, and therefore can only give rough predictions.

\section{Feasibility assessment and QKD test}
\label{sec:test}

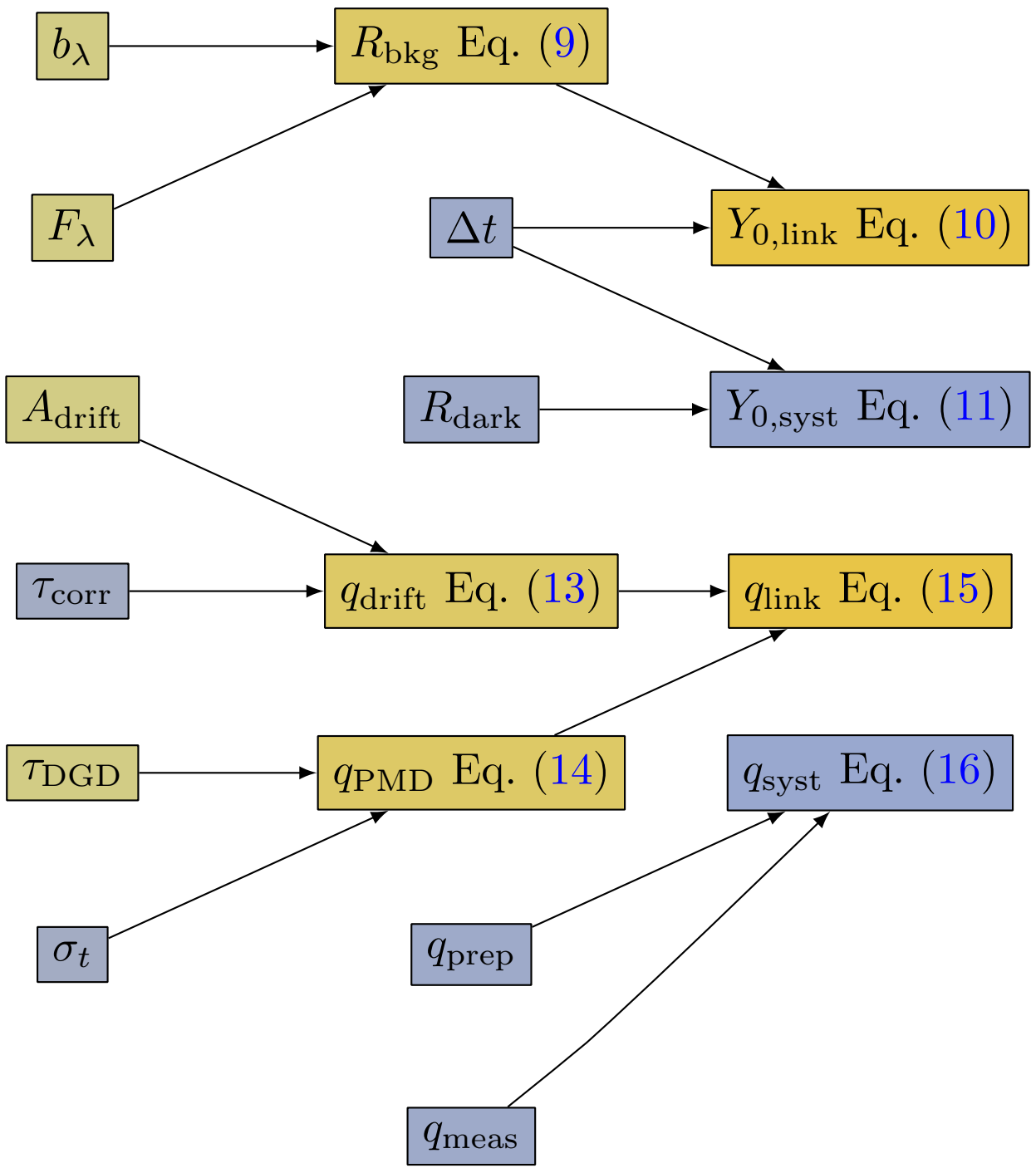
\begin{figure}
  \centering
  \begin{tikzpicture}[>=latex,line join=bevel,]
\begin{scope}
  \pgfsetstrokecolor{black}
  \definecolor{strokecol}{rgb}{1.0,1.0,1.0};
  \pgfsetstrokecolor{strokecol}
  \definecolor{fillcol}{rgb}{1.0,1.0,1.0};
  \pgfsetfillcolor{fillcol}
  \filldraw (0.0bp,0.0bp) -- (0.0bp,282.0bp) -- (234.0bp,282.0bp) -- (234.0bp,0.0bp) -- cycle;
\end{scope}
\begin{scope}
  \pgfsetstrokecolor{black}
  \definecolor{strokecol}{rgb}{1.0,1.0,1.0};
  \pgfsetstrokecolor{strokecol}
  \definecolor{fillcol}{rgb}{1.0,1.0,1.0};
  \pgfsetfillcolor{fillcol}
  \filldraw (0.0bp,0.0bp) -- (0.0bp,282.0bp) -- (234.0bp,282.0bp) -- (234.0bp,0.0bp) -- cycle;
\end{scope}
\begin{scope}
  \pgfsetstrokecolor{black}
  \definecolor{strokecol}{rgb}{1.0,1.0,1.0};
  \pgfsetstrokecolor{strokecol}
  \definecolor{fillcol}{rgb}{1.0,1.0,1.0};
  \pgfsetfillcolor{fillcol}
  \filldraw (0.0bp,0.0bp) -- (0.0bp,282.0bp) -- (234.0bp,282.0bp) -- (234.0bp,0.0bp) -- cycle;
\end{scope}
  \node (y0l) at (207.0bp,223.0bp) [draw,fill=gold1,rectangle] {$Y_{0, \mathrm{link}}$ Eq.~\eqref{eq:y0link}};
  \node (y0s) at (207.0bp,182.0bp) [draw,fill=mutedblue1,rectangle] {$Y_{0, \mathrm{syst}}$ Eq.~\eqref{eq:y0syst}};
  \node (ql) at (207.0bp,141.0bp) [draw,fill=gold1,rectangle] {$q_{\mathrm{link}}$ Eq.~\eqref{eq:qlink}};
  \node (qs) at (207.0bp,100.0bp) [draw,fill=mutedblue1,rectangle] {$q_{\mathrm{syst}}$ Eq.~\eqref{eq:qsyst}};
  \node (qdrift) at (117.0bp,141.0bp) [draw,fill=gold2,rectangle] {$q_{\mathrm{drift}}$ Eq.~\eqref{eq:qdrift}};
  \node (qpmd) at (117.0bp,100.0bp) [draw,fill=gold2,rectangle] {$q_{\mathrm{PMD}}$ Eq.~\eqref{eq:qpmd}};
  \node (qprep) at (117.0bp,59.0bp) [draw,fill=mutedblue2,rectangle] {$q_{\mathrm{prep}}$};
  \node (qmeas) at (117.0bp,18.0bp) [draw,fill=mutedblue2,rectangle] {$q_{\mathrm{meas}}$};
  \node (rbkg) at (117.0bp,264.0bp) [draw,fill=gold2,rectangle] {$R_{\mathrm{bkg}}$ Eq.~\eqref{eq:rbkg}};
  \node (deltat) at (117.0bp,223.0bp) [draw,fill=mutedblue2,rectangle] {$\Delta t$};
  \node (rdark) at (117.0bp,182.0bp) [draw,fill=mutedblue2,rectangle] {$R_{\mathrm{dark}}$};
  \node (adrift) at (27.0bp,182.0bp) [draw,fill=gold3,rectangle] {$A_{\mathrm{drift}}$};
  \node (taucorr) at (27.0bp,141.0bp) [draw,fill=mutedblue3,rectangle] {$\tau_{\mathrm{corr}}$};
  \node (taudgd) at (27.0bp,100.0bp) [draw,fill=gold3,rectangle] {$\tau_{\mathrm{DGD}}$};
  \node (sigmat) at (27.0bp,59.0bp) [draw,fill=mutedblue3,rectangle] {$\sigma_t$};
  \node (blambda) at (27.0bp,264.0bp) [draw,fill=gold3,rectangle] {$b_\lambda$};
  \node (flambda) at (27.0bp,223.0bp) [draw,fill=gold3,rectangle] {$F_\lambda$};
  \draw [->] (qdrift) ..controls (151.89bp,141.0bp) and (160.18bp,141.0bp)  .. (ql);
  \draw [->] (qpmd) ..controls (152.23bp,115.95bp) and (160.96bp,120.01bp)  .. (ql);
  \draw [->] (qprep) ..controls (152.23bp,74.95bp) and (160.96bp,79.014bp)  .. (qs);
  \draw [->] (qmeas) ..controls (141.32bp,37.596bp) and (142.68bp,38.81bp)  .. (144.0bp,40.0bp) .. controls (155.95bp,50.813bp) and (168.8bp,63.134bp)  .. (qs);
  \draw [->] (rbkg) ..controls (152.23bp,248.05bp) and (160.96bp,243.99bp)  .. (y0l);
  \draw [->] (deltat) ..controls (151.89bp,223.0bp) and (160.18bp,223.0bp)  .. (y0l);
  \draw [->] (deltat) ..controls (152.23bp,207.05bp) and (160.96bp,202.99bp)  .. (y0s);
  \draw [->] (rdark) ..controls (151.89bp,182.0bp) and (160.18bp,182.0bp)  .. (y0s);
  \draw [->] (adrift) ..controls (62.234bp,166.05bp) and (70.956bp,161.99bp)  .. (qdrift);
  \draw [->] (taucorr) ..controls (61.886bp,141.0bp) and (70.183bp,141.0bp)  .. (qdrift);
  \draw [->] (taudgd) ..controls (61.886bp,100.0bp) and (70.183bp,100.0bp)  .. (qpmd);
  \draw [->] (sigmat) ..controls (62.234bp,74.95bp) and (70.956bp,79.014bp)  .. (qpmd);
  \draw [->] (blambda) ..controls (61.886bp,264.0bp) and (70.183bp,264.0bp)  .. (rbkg);
  \draw [->] (flambda) ..controls (62.234bp,238.95bp) and (70.956bp,243.01bp)  .. (rbkg);
\end{tikzpicture}
  \caption{Scheme of the relations between physical parameters relevant for QKD performance, complementary to Fig.~\ref{fig:graph_analysis}.}
  \label{fig:graph_physical}
\end{figure}

The measured parameters of the links can be used to evaluate the feasibility of QKD deployments.
As illustrated by the diagram of Fig.~\ref{fig:graph_physical}, a full analysis requires cross-checking the measured parameters with the specifications of the QKD devices, to arrive at a prediction of the expected performance (see also Fig.~\ref{fig:graph_analysis}).
Even without such specifications, reasonable bounds for them that guarantee a positive $\SKR$ can still be determined.
This can help in the design and procurement phase, so users can ask for specific requirements and manufacturers can easily evaluate whether they can meet them.

To exemplify, let us consider again the L3--A link, for which we have reported a full characterization, and is representative of the typical lengths and losses of a long-distance QKD link.
We have measured $\eta_{\mathrm{link}} \approx \num{0.012}$ (Tab.~\ref{tab:lengths_attenuations}), $R_{\mathrm{bkg}} \approx \SI{0.9}{\kilo\hertz}$ (Tab.~\ref{tab:background}), and $q_{\mathrm{link}} \lesssim \num{1e-3}$ (Eq.~\eqref{eq:qlink}).
In our performance model, we need four system parameters: $\eta_{\mathrm{syst}}$, $R_{\mathrm{dark}}$, $\Delta t$ (for $Y_{0,\mathrm{syst}}$) and $q_{\mathrm{syst}}$.
Together with the measured link parameters $\eta_{\mathrm{link}}$, $R_{\mathrm{bkg}}$ and $q_{\mathrm{link}}$, these allow predicting the expected $\SKR$ as depicted by the diagrams of Figs.~\ref{fig:graph_analysis} and \ref{fig:graph_physical}.
Although an automated numerical search of the boundary of the region of positive $\SKR$ in the four-parameter space is possible, for the purpose of gaining more insights and providing general guidelines, we can more simply fix some parameters to reasonable values and find the boundary in the remaining subspace.
For instance, for an ideal system in which $\eta_{\mathrm{syst}}=\num{0}$, $R_{\mathrm{dark}}=\SI{0}{\hertz}$ and $q_{\mathrm{syst}}=\num{0}$, we can find that the maximum tolerable value of the temporal window is $\Delta t_{\max} \approx \SI{1.2}{\micro\second}$.
Modern QKD devices operate at much higher repetition rates than the inverse of $\Delta t_{\max}$, therefore this is an easily achievable bound \cite{Takesue2007,Dixon2009,Sasaki2011,Wang2012}.
Let us move to the opposite end of the spectrum and consider a realistic system that uses SPADs as detectors and a poorer discrimination of quantum states.
For $\eta_{\mathrm{syst}}=\num{0.1}$, $R_{\mathrm{dark}}=\SI{1}{\kilo\hertz}$ and $q_{\mathrm{syst}}=\num{0.02}$, we find $\Delta t_{\max} \approx \SI{60}{\nano\second}$, which is still achievable but more stringent.

These are good indications that the L3--A link is suitable for QKD deployment, assuming reasonable system parameters based on the state of the art.
For the latter set of realistic values and $\Delta t = \SI{5}{\nano\second}$, the L3--A link still supports $\SKR\approx \num{2e-4}$ bits per pulse, and even further reducing $\eta_{\mathrm{syst}}$ by \SI{10}{\dB} to account for phenomena that are neglected by the model still allows successful key generation with  $\SKR\approx\num{5e-6}$ bits per pulse.

\begin{figure}
  \centering
  \includegraphics[width=0.5\textwidth]{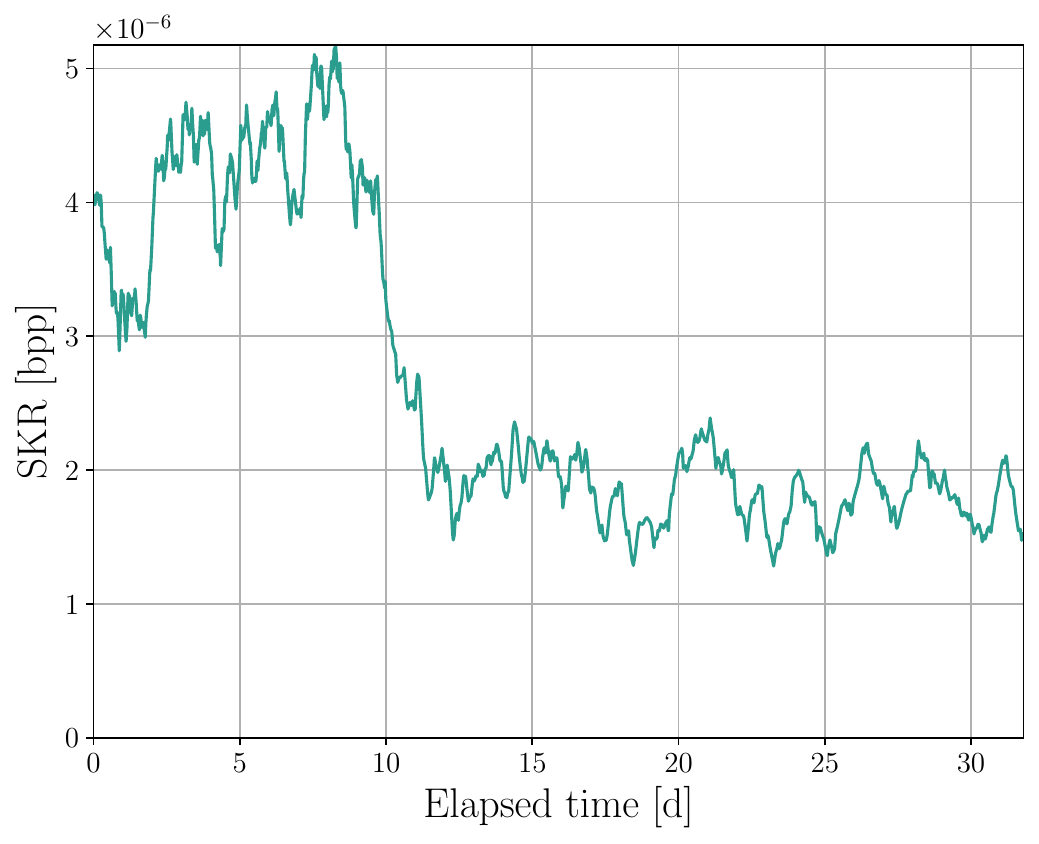}
  \caption{$\SKR$ over time for the L3--A link, plotted with a running mean of \SI{10}{\hour}.}
  \label{fig:qkd_plot}
\end{figure}

To validate this experimentally, we report a one-month long session of QKD operation in the L3--A link, using a commercial system (ThinkQuantum Quky Edu-Pro), which yielded a positive $\SKR$ for the whole duration of the test.
Figure \ref{fig:qkd_plot} shows the $\SKR$ over time, which is \num{2.4e-6} bits per pulse on average.
This is calculated by dividing the reported $\SKR$ in the unit time by the measured repetition rate of the transmitter.
The reduction in $\SKR$ after about eight days coincides with an increase of the QBER, which we speculate can be attributable to a realignment of the measurement bases.
This is an automated procedure that these systems perform when the QBER is too high, and it is not guaranteed to reach the optimal alignment every time.
Indeed, the performance seems to be limited by a relatively high QBER of \SI{4.2}{\percent} on the key basis, which we attribute either to $q_{\mathrm{syst}}$ or to the interplay between correction systems and link parameters that are not considered by the model.
The $\SKR$ and QBER values are affected by finite-size effects, limited sifting probability, imperfect efficiency of the error correction algorithm, detector imperfections such as dead time or afterpulsing, and sub-ideal choices of intensity levels dictated by engineering constraints, none of which are included in the simplified model.

\section{Conclusions}
\label{sec:conclusions}

We have reported here many of the physical parameters that are relevant for the deployment of QKD on a fiber network, and have shown how they can be used to predict the performance of a QKD system.
Our study, starting from the length of each link in the NQCIS network, touches the most important contributions to the secret key rate, namely the attenuation of the signal, the addition of background noise, and the deterioration of the quantum states due to the fiber channel.
We have suggested techniques to measure these parameters and ways to relate them to performance indicators, providing guidelines for planning and evaluating QKD deployments.

While specific to our configuration and choice of protocol, the methodology we have used is general and can be adapted to other QKD scenarios.
For this, we have used an essential model that pedagogically relates the physical parameters to the performance of decoy-state BB84.
Although other protocols change the precise relations between quantities, the general structure of the model remains similar, and the physical parameters we have measured are still relevant.

More work is needed to accurately predict the performance of a deployed QKD system, and it is possible that no analytical model will ever capture the full complexity of the problem.
For instance, the interplay between channel and stabilization system for polarization drift is non-trivial, and our simple model does not account for it.
Simulations and experimental tests with real systems can help to fill this gap and develop more accurate models.

In the meantime, our work is comprehensive enough to provide a good starting point for the imminent deployments in EuroQCI, which share many features with the NQCIS network.
This will help the evaluation of both fiber links and QKD devices, and will allow for a more informed planning of the network.

\section*{Acknowledgements}
\label{sec:acknowledgements}
The authors would like to thank Gemma Vall-llosera and Dushko Kuzmanovski at Ericsson AB for the use of the M1 link, Magnus Rådahl at Norrtälje Energi AB for the use of the L3 link, the LiU team led by Guilherme B.~Xavier for the use of the L4 link, and Jörgen Engvall at STOKAB for support in the procurement of the M1, M2, and L3 links.

\section*{Funding}
\label{sec:funding}
This work is supported by the EuroQCI programs DIGITAL-2022-QCI-02-DEPLOY-NATIONAL (project number 101113375 - NQCIS) and CEF-DIG-2024-EUROQCI-WORKS (project number 101249648 - NordicQCI), cofunded by the European union, Vinnova and the Wallenberg Centre for Quantum Technology (WACQT).

\section*{Author contributions}
\label{sec:contributions}
GF led all the technical, theoretical, and experimental efforts and wrote the manuscript.
MAY and SGN implemented and ran most of the experiments reported in Sec.~\ref{sec:noise}.
CBM implemented and ran the experiments reported in Sec.~\ref{sec:pmd}.
HK, DKW, and MAY contributed to the experiments reported in Secs.~\ref{sec:length}, \ref{sec:attenuation} and \ref{sec:drift}.
PL supported the later part of the tests, and implemented the measurement reported in Fig.~\ref{fig:att_vs_wl}.
KG is the principal investigator of the NQCIS project, provided overall guidance, and contributed to the manuscript.

\section*{Data availability}
\label{sec:data}
The data and the tools used to generate the results presented in this work are available online \cite{Dataset}.

\bibliographystyle{apsrev4-2}
\bibliography{bibliography}

\end{document}